\documentclass[aps,pra,notitlepage,reprint,floatfix,amsmath,amssymb]{revtex4-1}

\usepackage{newtxtext,newtxmath}
\usepackage{graphicx}
\usepackage{siunitx}
\usepackage{mathrsfs}
\usepackage{hyperref}

\newcommand{\T}[2]{\left\la \p T_{#1}(\pvec{#2} | \pvec{k} )/\p\Omega_t\right\ra_\textrm{incoh}}

\renewcommand{\vec}[1]{\mathbf{{#1}}}
\newcommand{\pvec}[1]{\mathbf{{#1}}_\parallel }
\newcommand{\vecUnit}[1]{\mathbf{\hat{#1}}}
\newcommand{\pvecUnit}[1]{\mathbf{\hat{#1}}_\parallel }

\newcommand{\imu}{i}
\newcommand{\dint}{\mathrm{d}}
\renewcommand{\Re}{\mathrm{Re}\,}
\renewcommand{\Im}{\mathrm{Im}\,}

\newcommand{\ppol}{p}
\newcommand{\spol}{s}

\newcommand{\etal}{\textit{et al.}~}

\newcommand{\nn}{\nonumber}
\newcommand{\zxp}{\zeta({\textbf x}_{\|})}
\newcommand{\zxpp}{\zeta({\textbf x}\, '\!\!_{\|} )}
\newcommand{\bkp}{{\mathbf k}_{\|}}
\newcommand{\bqp}{{\mathbf q}_{\|}}
\newcommand{\bpp}{{\mathbf p}_{\|}}
\newcommand{\qp}{q_{\|}}
\newcommand{\pp}{p_{\|}}
\newcommand{\kp}{k_{\|}}
\newcommand{\bxp}{{\textbf x}_{\|}}

\renewcommand{\a}{\alpha}
\newcommand{\p}{\partial }

\newcommand{\e}{\varepsilon}
\newcommand{\sfr}{^{\frac{1}{2}}}

\newcommand{\w}{\omega }

\newcommand{\la}{\langle}
\newcommand{\ra}{\rangle}
\newcommand{\xp}{x_{\|}}

\begin{document}


\title{Numerical studies of the transmission of light through a two-dimensional randomly rough interface}
\author{\O.S. Hetland$^1$}
\email{oyvind.hetland@ntnu.no}
\author{A.A. Maradudin$^2$}
\author{T. Nordam$^1$}
\author{P.A. Letnes$^1$}
\author{I. Simonsen$^{1,3}$}
\affiliation{$^1$Department of Physics, NTNU Norwegian University of Science and Technology, NO-7491 Trondheim, Norway}
\affiliation{$^2$Department of Physics and Astronomy, University of California, Irvine CA 92697, U.S.A.}
\affiliation{$^3$Surface du Verre et Interfaces, UMR 125 CNRS/Saint-Gobain, F-93303 Aubervilliers, France}

\date{December 13, 2016}

\begin{abstract}
  The transmission of polarized light through a two-dimensional randomly rough interface between two dielectric media has been much less studied, by any approach, than the reflection of light from such an interface.  We have derived a reduced Rayleigh equation for the transmission amplitudes when p- or s-polarized light is incident on this type of interface, and have obtained rigorous, purely numerical, nonperturbative solutions of it.
  The solutions are used to calculate the transmissivity and transmittance of the interface, the mean differential transmission coefficient, and the full angular distribution of the intensity of the transmitted light.
  These results are obtained for both the case where the medium of incidence is the optically less dense medium and in the case where it is the optically more dense medium.
  Optical analogues of Yoneda peaks observed in the scattering of x-rays from metallic and non-metallic surfaces are present in the results obtained in the former case.
  For p-polarized incident light we observe Brewster scattering angles, angles at which the diffuse transmitted intensity is zero in a single-scattering approximation, which depend on the angle of incidence in contrast to the Brewster angle for flat-surface reflection.
 \end{abstract}

\maketitle


\section{Introduction}
In the theoretical and experimental studies of the interaction of an electromagnetic wave with a two-dimensional randomly rough dielectric surface, the great majority have been devoted to the reflection problem~\cite{Simonsen2011,Nordam2013a,Leskova2011}, and less attention has been paid to studies of the transmission of light through such surfaces.
Greffet~\cite{Greffet1988} obtained a reduced Rayleigh equation for the transmission amplitudes in the case where light incident from vacuum is transmitted through a two-dimensional randomly rough interface into a dielectric medium, and obtained a recursion relation for the successive terms in the expansions of the amplitudes in powers of the surface profile function.
Kawanishi~\etal\cite{Kawanishi1997}, by the use of the stochastic functional approach, studied the case where a two-dimensional randomly rough interface between two dielectric media is illuminated by p- and s-polarized light from either medium.
Properties of the light transmitted through, as well as reflected from, the interface were examined. This theoretical approach is perturbative in nature and can be applied only to weakly rough surfaces.
Nevertheless, Kawanishi~\etal obtained several interesting properties of the transmitted light that are associated with the phenomenon of total internal reflection when the medium of transmission is the optically denser medium.
These include the appearance of Yoneda peaks in the intensity of the transmitted light as  a function of the angle of transmission for a fixed value of the angle of incidence.
Yoneda peaks are sharp asymmetric peaks at
the critical polar angle of transmission for which the wavenumber of incidence turns non-propagating when the medium of transmission is the optically more dense medium.
Although well known in the scattering of x-rays from both metallic~\cite{Yoneda1963,Sinha1988,Gorodnichev1988,Renaud2009} and non-metallic~\cite{Dosch1987,Stepanov2000,Kitahara2002,Gasse2016} surfaces, the paper by Kawanishi~\etal apparently marks their first explicit appearance in optics.
Yoneda peaks were recently observed experimentally for a configuration of reflection from a randomly rough dielectric interface, when the medium of incidence was the optically denser medium \cite{Gonzalez-Alcalde2016}. The physical origin of the Yoneda peak phenomenon is not clear \cite{Hetland2016a}.

For p-polarized incident light Kawanishi~\etal also observed angles of zero scattering intensity, to first order in their approach, in the distributions of the intensity of the incoherently reflected and transmitted light.
Due to their resemblance to the Brewster angle in the reflectivity from a flat interface, they dubbed these angles the ``Brewster scattering angles''.
These were observed, in both reflection and transmission, for light incident from either medium, and were found to be strongly dependent on the angle of incidence.
The Brewster scattering angles can be observed to be part of the mechanisms that result in a strong dependence on polarization in the scattering distributions of incoherently scattered light.
Nieto-Vesperinas and S\'{a}nchez-Gil~\cite{Nieto-Vesperinas1992} observed this strong dependence on polarization in their numerical investigations of incoherent transmission through one-dimensional dielectric surfaces, but they did not investigate this dependence any further.

Soubret~\etal\cite{Soubret2001a} also obtained a reduced Rayleigh equation for the transmission amplitudes in the case where light incident from one dielectric medium is transmitted into a second dielectric medium through a two-dimensional randomly rough interface.  However, only perturbative solutions of this equation were obtained by them, and only for vacuum as the medium of incidence.

In this paper we present a theoretical study of the transmission of light through a two-dimensional randomly rough interface between two dielectric  media, free from some of the limitations and approximations present in the earlier studies of this problem.
We obtain a reduced Rayleigh equation for the transmission amplitudes in the case where light incident from a dielectric medium whose dielectric constant is $\e_1$ is transmitted through a two-dimensional randomly rough interface into a dielectric medium whose dielectric constant is $\e_2$.  The dielectric constant $\e_1$ can be larger or smaller than the dielectric constant $\e_2$.
Thus, effects associated with total internal reflection are included in the solutions of this equation.  Instead of solving the reduced Rayleigh equation as an expansion in powers of the surface profile function, in this work we obtain a rigorous, purely numerical, nonperturbative solution of it.
This approach enables us to calculate the transmissivity and transmittance of the system studied, the in-plane co- and cross-polarized, and the out-of-plane co- and cross-polarized incoherent (diffuse) scattering contributions to the mean differential transmission coefficient, and the full angular dependence of the total scattered intensity, all in a nonperturbative fashion.

Numerical studies of similar systems and phenomena, obtained through a corresponding numerical method but in reflection, have previously been reported in Refs.~\citenum{Gonzalez-Alcalde2016} and ~\citenum{Hetland2016a}. Both Yoneda peaks and Brewster scattering angles were reported and discussed in-depth in Ref.~\citenum{Hetland2016a}, and an experimental observation of Yoneda peaks were presented in Ref.~\citenum{Gonzalez-Alcalde2016}.
As such, the currently presented work serves to add to the fuller understanding of the scattering behaviour of randomly rough dielectric interfaces.

\section{Scattering System}
\begin{figure}
  \centering
  \includegraphics{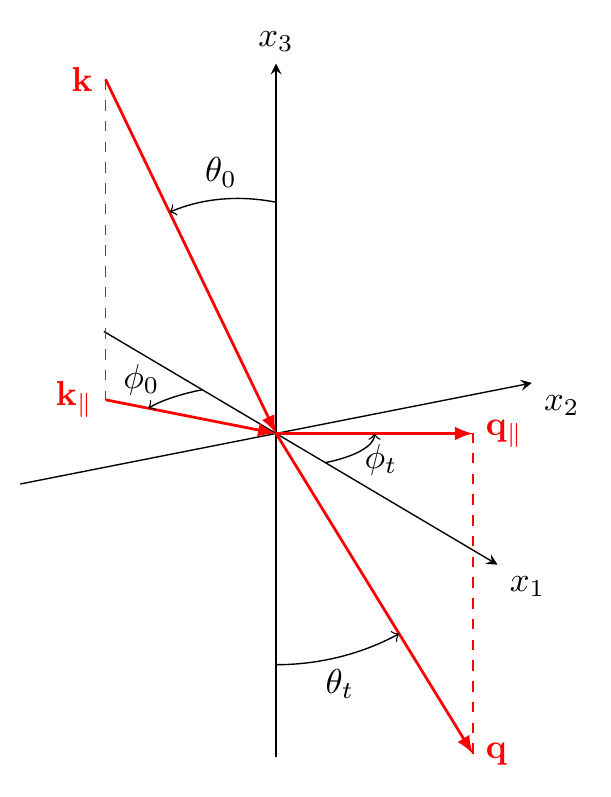} 
  \caption{A sketch of the scattering geometry assumed in this work. The figure also shows the coordinate system used, angles of incidence ($\theta_0,\phi_0$) and transmission ($\theta_t,\phi_t$), and the corresponding lateral wave vectors $\pvec{k}$ and $\pvec{q}$.}
  \label{fig:scattering_geometry}
\end{figure}

The system we study in this paper consists of a dielectric medium (medium~1), whose dielectric constant is $\e_1$, in the region $x_3 > \zxp$, and a dielectric medium (medium~2), whose dielectric constant is $\e_2$, in the region $x_3 < \zxp$ [Fig.~\ref{fig:scattering_geometry}].  Here $\bxp = (x_1, x_2, 0)$ is an arbitrary vector in the plane $x_3 = 0$, and we assume that both $\e_1$ and $\e_2$ are real and positive.

The surface profile function $\zxp$ is assumed to be a single-valued function of $\bxp$ that is differentiable with respect to $x_1$ and $x_2$, and constitutes a stationary, zero-mean, isotropic, Gaussian random process defined by
\begin{align}
  \la \zxp \zxpp \ra = \delta^2W(|\bxp - \bxp ' |) , \label{eq:2.1}
\end{align}
where $W(x_\parallel)$ is the normalized surface height autocorrelation function, with the property that $W(0)=1$.
The angle brackets here and in all that follows denote an average over the ensemble of realizations of the surface profile function.  The root-mean-square height of the surface is given by
\begin{align}
  \delta = \la \zeta^2(\bxp )\ra^{\frac{1}{2}} .\label{eq:2.2}
\end{align}
The power spectrum of the surface roughness $g(\kp )$ is defined by
\begin{align}
  g(\kp ) = \int \!\dint^2\xp \; W(\xp )\exp (-\imu\bkp \cdot \bxp ) , \label{eq:2.3}
\end{align}
where $\bkp = (k_1,k_2,0)$ is a lateral wave vector, $\kp=\left|\pvec{k}\right|$ and $\xp=\left|\bxp\right|$.
We will assume for the normalized surface height autocorrelation function $W(\xp )$ the Gaussian function
\begin{align}
  W(\xp ) = \exp \left(- \frac{x_\parallel^2}{a^2} \right) , \label{eq:2.4}
\end{align}
where the characteristic length $a$ is the transverse correlation length of the surface roughness.  The corresponding power spectrum is given by
\begin{align}
  g(\kp ) = \pi a^2 \exp \left(- \frac{\kp^2 a^2}{4} \right) . \label{eq:2.5}
\end{align}

\section{The Reduced Rayleigh Equation}
The interface $x_3 = \zxp$ is illuminated from the region $x_3 > \zxp$ (medium 1) by an electromagnetic wave of frequency $\w$.  The total electric field in this region is the sum of an incoming incident field and an outgoing scattered field,
\begin{align}
\begin{aligned}
  \vec{E}^>(\vec{x} |\omega ) = \; &\vec{E}_0(\pvec{k}) \exp [\imu\vec{Q}_0(\pvec{k}) \cdot \vec{x} ]
  \\
  &+ \int \!\frac{\dint^2\qp}{(2\pi )^2}\; \vec{A}(\pvec{q} ) \exp [\imu\vec{Q}_1(\pvec{q} )\cdot \vec{x} ],
  \label{eq:3.1}
\end{aligned}
\end{align}
while the electric field in the region $x_3 < \zxp$ is an outgoing transmitted field,
\begin{align}
  \vec{E}^<(\vec{x} |\omega ) = \int \!\frac{\dint^2\qp}{(2\pi )^2}\; \vec{B}(\pvec{q} ) \exp [\imu \vec{Q}_2(\pvec{q} )\cdot \vec{x} ]. \label{eq:3.2}
\end{align}
In writing these equations we have introduced the functions
\begin{subequations}
  \label{eq:3.3}
  \begin{align}
    \vec{Q}_0(\bkp ) &= \pvec{k} - \alpha_1(\kp ) \vecUnit{x}_3     \label{eq:3.3a}\\
    \vec{Q}_1(\bqp ) &= \pvec{q} + \alpha_1(\qp )  \vecUnit{x}_3    \label{eq:3.3b}\\
    \vec{Q}_2(\bqp ) &= \pvec{q} - \alpha_2(\qp )  \vecUnit{x}_3 ,  \label{eq:3.3c}
  \end{align}
\end{subequations}
where $(i = 1, 2)$
\begin{align}
  \alpha_i(\qp )
  &=
  \begin{cases}
    \sqrt{ \e_i\left(\frac{\w}{c}\right)^2 - \qp^2}, &  \qp \leq \sqrt{\e_i}\,\w/c,
    \\*[0.4em]
    \imu \sqrt{ \qp^2 - \e_i\left(\frac{\w}{c}\right)^2 }, &  \qp > \sqrt{\e_i}\,\w/c.
  \end{cases}
  \label{eq:3.4}
\end{align}
Here $\bqp = (q_1,q_2,0)$, $\qp=\left|\pvec{q}\right|$, and a caret over a vector indicates that it is a unit vector.  A time dependence of the field of the form $\exp (-\imu\w t)$ has been assumed, but not indicated explicitly.

The boundary conditions satisfied  by these fields at the interface $x_3 = \zxp$ are the continuity of the tangential components of the electric field:
\begin{align}
  \vec{n} &\times \vec{E}_0(\bkp )  \exp [  \imu \pvec{k} \cdot \bxp -  \imu \alpha_1(\kp ) \zxp ]
  \nn\\
  &\quad + \int \frac{\dint^2\qp}{(2\pi )^2}\; \vec{n} \times \vec{A}(\bqp ) \exp [\imu\bqp \cdot\bxp + \imu\alpha_1(\qp )\zxp ]
  \nn\\
  &= \int \frac{\dint^2\qp}{(2\pi )^2}\; \vec{n}\times \vec{B} (\bqp ) \exp [\imu\bqp \cdot\bxp - \imu\alpha_2(\qp )\zxp ] ;
   \label{eq:3.5}
\end{align}
the continuity of the tangential components of the magnetic field:
\begin{align}
\begin{aligned}[b]
  \vec{n} \:\times &\: [\imu\vec{Q}_0(\bkp )  \times \vec{E}_0(\bkp ) ] \exp [\imu\bkp \cdot\bxp - \imu\alpha_1(\kp ) \zxp ]
  \\
  &+ \int \frac{\dint^2\qp}{(2\pi )^2}\; \vec{n} \times [\imu\vec{Q}_1(\bqp ) \times \vec{A}(\bqp )]
  \\
  &\times \exp [\imu\bqp\cdot\bxp + \imu\alpha_1(\qp )\zxp ]
  \\
  = &\int \frac{\dint^2\qp}{(2\pi )^2}\; \vec{n} \times [\imu\vec{Q}_2(\bqp ) \times \vec{B}(\bqp )]
  \\
  &\times \exp [\imu\bqp\cdot\bxp - \imu\alpha_2(\qp )\zxp ] ;
    \label{eq:3.6}
\end{aligned}
\end{align}
and the continuity of the normal component of the electric displacement:
\begin{align}
 \e_1 \vec{n}\: \cdot\:  &\vec{E}_0(\bkp )  \exp [\imu\bkp \cdot\bxp - \imu\alpha_1(\kp )\zxp ]
 \nn\\
 &+ \e_1\int \!\frac{\dint^2\qp}{(2\pi )^2}\; \vec{n} \cdot \vec{A}(\bqp ) \exp [\imu\bqp \cdot \bxp + \imu\alpha_1(\qp ) \zxp ]
 \nn\\
 =\; \e_2 &\int \!\frac{\dint^2 \qp}{(2\pi )^2}\; \vec{n} \cdot \vec{B} (\bqp ) \exp [\imu\bqp \cdot \bxp - \imu\alpha_2 (\qp ) \zxp ].\label{eq:3.7}
\end{align}
The vector $\vec{n}\equiv \vec{n}(\pvec{x})$ entering these equations is a vector normal to the surface $x_3 = \zxp$ at each point of it, directed into medium $1$:
\begin{align}
\begin{aligned}
  \label{eq:3.8}
    \vec{n}(\pvec{x}) = \left(-\frac{\p \zxp}{\p x_1}, - \frac{\p\zxp}{\p x_2}, 1\right).
\end{aligned}
\end{align}
%
Strictly speaking the continuity of the tangential components of the electric and magnetic fields across the interface, Eqs.~\eqref{eq:3.5} and \eqref{eq:3.6}, are sufficient (and necessary) boundary conditions on  electromagnetic fields~\cite{Yeh1993}. Hence, the continuity of the normal components of the electric displacement [Eq.~(\ref{eq:3.7})] and the magnetic induction are redundant. However, the inclusion of Eq.~(\ref{eq:3.7}) enables us to eliminate the scattering amplitude ${\textbf A}(\bqp )$ from consideration, and thus to obtain an equation that relates the transmission amplitude $\vec{B}(\bqp )$ to the amplitude of the incident field ${\textbf E}_0(\bkp )$. This we do in the following manner.

We take the vector cross product of Eq.~(\ref{eq:3.5}) with $\e_1 \vec{Q}_0(\bpp ) \exp [-\imu\bpp \cdot\bxp + \imu\alpha_1(\pp )\zxp ] $; then multiply Eq.~(\ref{eq:3.6}) by $-\imu\e_1\exp [-\imu\pvec{p}\cdot\bxp $ $+\imu\alpha_1(\pp ) \zxp ]$; and finally multiply Eq.~(\ref{eq:3.7}) by $-\vec{Q}_0(\bpp ) \exp [-\imu \bpp \cdot \bxp + \imu\alpha_1(\pp ) \zxp ]$, where $\bpp$ is an arbitrary wave vector in the plane $x_3 = 0$.  When we add the three equations obtained in this way, and integrate the sum over $\bxp$ we obtain an equation that can be written in the form
\begin{widetext}
\begin{align}
   \e_1 &
     \left\{
           \vec{Q}_0(\bpp )             \times \left[ \vec{V}_{E}(\bpp |\bkp ) \times \vec{E}_0(\bkp ) \right]
         + \vec{V}_{E}(\bpp |\bkp )      \times \left[ \vec{Q}_0(\bkp ) \times \vec{E}_0(\bkp )        \right]
         - \vec{Q}_0(\bpp )                     \left[ \vec{V}_{E}(\bpp |\bkp )\cdot \vec{E}_0(\bkp )  \right]
     \right\}
  \nn\\ &
  \quad
  + \e_1\int \!\frac{\dint^2\qp}{(2\pi )^2}\;
       \Big\{
             \vec{Q}_0(\bpp )       \times \left[ \vec{V}_{A}(\bpp |\bqp )\times \vec{A}(\bqp ) \right]
           + \vec{V}_A(\bpp |\bqp ) \times \left[ \vec{Q}_1(\bqp )        \times \vec{A}(\bqp ) \right]
           - \vec{Q}_0(\bpp )              \left[ \vec{V}_{A}(\bpp |\bqp )\cdot \vec{A}(\bqp )  \right]
       \Big\}
  \nn\\&
   = \int \frac{\dint^2\qp}{(2\pi )^2}\,
     \Big\{
           \e_1 \vec{Q}_0(\bpp )       \times \left[ \vec{V}_B(\bpp |\bqp )   \times \vec{B}(\bqp )   \right]
         + \e_1 \vec{V}_B(\bpp |\bqp ) \times \left[ \vec{Q}_2(\bqp )         \times \vec{B}(\bqp )   \right]
         - \e_2 \vec{Q}_0 (\bpp )             \left[ \vec{V}_{B}(\bpp |\bqp ) \cdot \vec{B}(\bqp )    \right]
      \Big\},
      \label{eq:3.9}
  %
\end{align}
\end{widetext}

where we define
\begin{subequations}
\label{eq:3.10}
\begin{align}
  \vec{V}_E(\bpp |\bkp )
  &= \vec{V}\left( -\alpha_1(\pp ) + \alpha_1(\kp ) | \bpp - \bkp   \right)
  \label{eq:3.10a}
  \\
  \vec{V}_A(\bpp |\bqp )
    &= \vec{V}\left( -\alpha_1(\pp )-\alpha_1(\qp ) | \bpp - \bqp   \right)
   \label{eq:3.10b}
   \\
   \vec{V}_B(\bpp |\bqp )
   &= \vec{V}\left( -\alpha_1(\pp ) + \alpha_2(\qp ) | \bpp -\bqp   \right),
   \label{eq:3.10c}
\end{align}
\end{subequations}
with
\begin{subequations}
  \label{eq:3.11-is}
  \begin{align}
    \vec{V}( \gamma  | \pvec{Q} )
    &=
    \int \!\dint^2x_\parallel\;
    \vec{n}(\pvec{x})
    \exp \left(-\imu \pvec{Q} \cdot \pvec{x}  \right)
    \exp \left[ - \imu \gamma \zeta(\pvec{x}) \right].
    \label{eq:3.11a-is}
  \end{align}
  It is shown in Appendix~\ref{app:Details} that
  \begin{align}
    \vec{V}( \gamma  | \pvec{Q} )
    &=
    \frac{ I(\gamma | \pvec{Q} ) }{ \gamma }
    \left( \pvec{Q} + \gamma \vecUnit{x}_3  \right)
    - \left( 2 \pi \right)^2 \delta\left( \pvec{Q} \right) \frac{ \pvec{Q} }{ \gamma },
    \label{eq:3.11b-is}
  \end{align}
\end{subequations}
where
\begin{align}
  I( \gamma |  \pvec{Q} )
  &=
  \int \!\dint^2x_\parallel\;
     \exp \left(-\imu \pvec{Q} \cdot \pvec{x}  \right)
  \exp \left[ - \imu \gamma \zeta(\pvec{x}) \right].
  \label{eq:3.12-is}
\end{align}

When Eqs.~(\ref{eq:3.10})  and \eqref{eq:3.11-is} are substituted into Eq.~(\ref{eq:3.9}), the latter becomes
%
\begin{align}
 (2&\pi )^2 \delta (\bpp - \bkp ) 2 \e_1 \frac{\bkp \cdot (\bpp - \bkp )}{-\alpha_1(\pp )+\alpha_1(\kp )} \vec{E}_0(\bkp )
 \nn&\\
 =\; &(\e_1-\e_2) \int \frac{\dint^2\qp}{(2\pi )^2}\; \frac{I(-\alpha_1(\pp )+\alpha_2(\qp )|\bpp - \bqp )}{-\alpha_1(\pp )+\alpha_2(\qp )}
 \nn\\
 & \times
  \left\{ - \e_1 \left(\frac{\omega}{c}\right)^2 \vec{B}(\bqp ) + \left[ \vec{Q}_0(\bpp )\cdot \vec{B}(\bqp ) \right] \vec{Q}_0(\bpp )\right\}.\label{eq:3.16}%
\end{align}
In obtaining this result we have used the result that the singular term of $\vec{V}_B(\bpp |\bqp )$ does not contribute to the right hand side of Eq.~\eqref{eq:3.9}, since $\bpp = \bqp$ leaves $-\alpha_1(\pp )+\alpha_2(\qp )$ nonzero (see Appendix A). If we note that
\begin{align}
  -\alpha_1(\pp ) + \alpha_1(\kp ) = \frac{\bkp \cdot (\bpp -\bkp)}{\alpha_1(\kp )} + \mathcal{O} \left((\bpp - \bkp )^2 \right) ,
  \label{eq:3.17}
\end{align}
the left-hand side of Eq.~(\ref{eq:3.16}) becomes $(2\pi )^2\delta (\bpp -\bkp )2\e_1\alpha_1(\kp ){\textbf E}_0(\bkp )$.  Thus we have an equation for the transmission amplitude ${\textbf B}(\bqp )$ alone:
\begin{align}
   &\int \frac{\dint^2\qp}{(2\pi )^2} \frac{I(-\alpha_1(\pp )+\alpha_2(\qp )|\pvec{p} -\pvec{q} )}{-\alpha_1(\pp ) + \alpha_2(\qp )}
   \nn\\
  &\times \left\{ - \e_1\left(\frac{\omega}{c}\right)^2  \vec{B}(\bqp ) + \left[ \vec{Q}_0(\bpp )\cdot\vec{B}(\bqp )\right] \vec{Q}_0(\bpp )\right\}
  \nn\\
   = (&2\pi )^2 \delta (\bpp - \bkp ) \frac{ 2\e_1\alpha_1(\kp ) }{ \e_1 - \e_2 }  \vec{E}_0(\bkp ) .
%
  \label{eq:3.18}
\end{align}

We now write the vectors $\vec{E}_0(\bkp )$ and $\vec{B}(\bqp )$ in the forms
\begin{subequations}
  \label{eq:3.19}
  \begin{align}
    \vec{E}_0(\bkp ) = \vec{\hat{e}}_p^{(i)}(\bkp ) E_{0p}(\bkp ) + \vecUnit{e}_s^{(i)}(\bkp ) E_{0s}(\bkp ) ,
    \label{eq:3.19a}
  \end{align}
  where
  \begin{align}
    \vecUnit{e}^{(i)}_p(\bkp ) &= \frac{c}{\sqrt{\e_1}\w} \left[ \pvecUnit{k}\alpha_1(\kp ) + \vecUnit{x}_3\kp \right]
    \label{eq:3.19b}\\
    \vecUnit{e}^{(i)}_s(\bkp ) &= \vecUnit{x}_3\times \pvecUnit{k},
    \label{eq:3.19c}
  \end{align}
\end{subequations}
and
\begin{subequations}
  \label{eq:3.20}
  \begin{align}
    \vec{B}(\bqp ) = \vecUnit{e}^{(t)}_p(\bqp ) B_p(\bqp ) + \vecUnit{e}^{(t)}_{s}(\bqp )B_s(\bqp ) ,
    \label{eq:3.20a}
  \end{align}
  where
  \begin{align}
    \vecUnit{e}^{(t)}_p(\bqp ) &= \frac{c}{\sqrt{\e_2}\w} \left[  \pvecUnit{q} \alpha_2(\qp ) + \vecUnit{x}_3\qp \right],
  \label{eq:3.20b}\\
    \vecUnit{e}^{(t)}_s(\bqp ) &= \vecUnit{x}_3 \times \pvecUnit{q} .
    \label{eq:3.20c}
  \end{align}
\end{subequations}
In these expressions $E_{0p}(\bkp )$ and $E_{0s}(\bkp )$ are the amplitudes of the p- and s-polarized components of the incident field with respect to the plane of incidence, defined by the vectors $\pvecUnit{k}$ and $\vecUnit{x}_3$.  Similarly, $B_p(\bqp )$ and $B_s(\bqp )$ are the amplitudes of the p- and s-polarized components of the transmitted field with respect to the plane of transmission  defined by the vectors $\pvecUnit{q}$ and $\vecUnit{x}_3$.

Our goal is to express $B_p(\bqp )$ and $B_s(\bqp )$ in terms of $E_{0p}(\bkp )$ and $E_{0s}(\bkp )$.  To this end we introduce three mutually  perpendicular  unit vectors:
\begin{subequations}
  \label{eq:3.21}
  \begin{align}
    \vecUnit{a}_0(\bpp ) &=  \frac{c}{\sqrt{\e_1}\w} \left[ \pvecUnit{p} - \vecUnit{x}_3\alpha_1(\pp ) \right]
    \label{eq:3.21a}\\
    \vecUnit{a}_1(\bpp ) &= \frac{c}{\sqrt{\e_1}\w} \left[ \pvecUnit{p}\alpha_1(\pp ) + \vecUnit{x}_3\pp \right]
    \label{eq:3.21b}\\
    \vecUnit{a}_2(\bpp ) &= \vecUnit{x}_3 \times \pvecUnit{p} .
    \label{eq:3.21c}
  \end{align}
\end{subequations}
We now take the scalar product of Eq.~(\ref{eq:3.18}) with each of these three unit vectors in turn, after $\vec{E}_0(\bkp )$ and $\vec{B}(\bqp )$ have been replaced by the right-hand sides of Eq.~(\ref{eq:3.19a}) and (\ref{eq:3.20a}), respectively.  The results are:
\begin{widetext}
\begin{subequations}
  \label{eq:3.22}
  \begin{align}
    & \vecUnit{a}_0(\bpp ) \cdot \mbox{\textrm Eq.~\protect\eqref{eq:3.18}}: \qquad 0 = 0 ;
    \label{eq:3.22a}\\
    %
    %
    & \vecUnit{a}_1(\bpp ) \cdot \mbox{\textrm Eq.~\protect\eqref{eq:3.18}}: \qquad
    \nn\\
    &\qquad
    \int \frac{\dint^2\qp}{(2\pi )^2}\; \frac{I(-\alpha_1(\pp )+\alpha_2(\qp )|\bpp - \bqp )}{-\alpha_1(\pp )+\alpha_2(\qp )}
    \bigg\{ - \sqrt{\frac{\e_1}{\e_2}} \left[\alpha_1(\pp )\, \pvecUnit{p} \cdot \pvecUnit{q}\,\alpha_2(\qp ) + \pp\qp \right] B_p(\bqp )
     \nn\\& \hspace{0.25\columnwidth}
    + \sqrt{\e_1}\frac{\w}{c} \alpha_1(\pp )\, \left[ \pvecUnit{p}\times \pvecUnit{q}\right]_3 \, B_s(\bqp ) \bigg\}
    =(2\pi )^2\delta (\bpp - \bkp ) \frac{2\e_1\alpha_1(\kp )}{\e_1-\e_2}  E_{0p}(\bkp );
    \label{eq:3.22b}\\
    %
    & \vecUnit{a}_2(\bpp ) \cdot \mbox{\textrm Eq.~\protect\eqref{eq:3.18}}:
    \nn\\
    & \qquad
    \int \frac{\dint^2\qp}{(2\pi )^2} \; \frac{I(-\alpha_1(\pp )+\alpha_2(\qp )|\bpp - \bqp )}{-\alpha_1(\pp ) + \alpha_2(\qp )}
    \bigg\{ - \frac{\e_1}{\sqrt{\e_2}} \frac{\w}{c} \, \left[ \pvecUnit{p} \times \pvecUnit{q} \right]_3 \,\alpha_2(\qp ) B_p(\bqp )
    - \e_1 \frac{\w^2}{c^2} \pvecUnit{p} \cdot \pvecUnit{q}  B_s(\bqp ) \bigg\}
    \nn\\& \hspace{0.25\columnwidth}
    = (2\pi  )^2 \delta (\bpp - \bkp ) \frac{ 2\e_1\alpha_1(\kp ) }{ \e_1-\e_2 }  E_{0s}(\bkp ).
    \label{eq:3.22c}
  \end{align}
\end{subequations}

These equations represent linear relations between $B_{p,s}(\bqp )$ and $E_{0p,s}(\bkp )$ which we write in the form $(\alpha = p, s, \, \beta = p, s)$
\begin{align}
  B_{\alpha}(\bqp ) = \sum_{\beta} T_{\alpha\beta}(\bqp |\bkp ) E_{0\beta}(\bkp ) .
  \label{eq:3.23}
\end{align}
On combining Eqs.~\eqref{eq:3.22} and \eqref{eq:3.23} we find that the transmission amplitudes $\{ T_{\alpha\beta}(\bqp |\bkp )\}$ are the solutions of the equation
\begin{align}
  \int \frac{\dint^2\qp}{(2\pi )^2} \; \frac{I(-\alpha_1(\pp )+\alpha_2(\qp )|\bpp -\bqp )}{-\alpha_1(\pp )+\alpha_2(\qp )}
  \vec{M} (\bpp |\bqp ) \vec{T}(\bqp |\bkp )
  = (2\pi )^2 \delta (\bpp - \bkp ) \frac{2\alpha_1(\kp )}{\e_2-\e_1}   \vec{I}_2 , \label{eq:3.24}
\end{align}
where
\begin{subequations}
  \label{eq:3.25}
  \begin{align}
    \vec{M}(\bpp |\bqp ) &=
    \left(
      \begin{array}{cc}
        \frac{1}{\sqrt{\e_1\e_2}} [\alpha_1(\pp )\, \pvecUnit{p} \cdot \pvecUnit{q}  \,\alpha_2(\qp ) + \pp\qp]
        & \quad
        - \frac{1}{\sqrt{\e_1}}\frac{\w}{c}\alpha_1(\pp )\, [\vec{\hat{p}}_{\|} \times \vec{\hat{q}}_{\|}]_3
        \\ 
        \frac{1}{\sqrt{\e_2}}\frac{\w}{c} [\vec{\hat{p}}_{\|}\times \vec{\hat{q}}_{\|}]_3 \,\alpha_2(\qp )
        & \quad
        \frac{\w^2}{c^2}\, \pvecUnit{p} \cdot \pvecUnit{q}
      \end{array}
    \right), \label{eq:3.25a}
    \\
     \vec{T} (\bqp |\bkp )
    &=
    \left(
      \begin{array}{cc}
        T_{pp}(\bqp |\bkp ) & T_{ps}(\bqp |\bkp )\\
        T_{sp}(\bqp |\bkp ) & T_{ss}(\bqp |\bkp )
      \end{array}
    \right),
    \label{eq:3.25b}
\end{align}
and
\begin{align}
  \vec{I}_2 &=
  \left(
    \begin{array}{cc}
      1 \; & 0\\
      0 \; & 1
    \end{array}
  \right) .
  \label{eq:3.25c}
\end{align}
\end{subequations}
Equation (\ref{eq:3.24}) is the reduced Rayleigh equation for the transmission amplitudes.

\section{The Mean Differential Transmission Coefficient}
The differential transmission coefficient $\p T/\p\Omega_t$ is defined such that $(\p T/\p\Omega_t)\dint\Omega_t$ is the fraction of the total time-averaged flux incident on the interface that is transmitted into the element of solid angle $\dint\Omega_t$ about the direction of transmission $(\theta_t,\phi_t)$.
To obtain the mean differential transmission coefficient we first note that the magnitude of the total time-averaged flux incident on the interface is given by
\begin{align}
  P_\textrm{inc} &= - \Re \frac{c}{8\pi} \int \!\dint^2\xp \left\{ \vec{E}^*_0(\bkp ) \times \left[\frac{c}{\w} \vec{Q}_0(\pvec{k} )\times \vec{E}_0 (\bkp ) \right] \right\}_3
  \exp \left\{ [-\imu \vec{Q}_0^*(\bkp ) + \imu \vec{Q}_0(\bkp ) ] \cdot \vec{x}  \right\}
  \nn\\
  &= -\Re \frac{c^2}{8\pi \w} \int \!\dint^2\xp\; \left\{ \left| \vec{E}_0 (\bkp ) \right|^2 \vec{Q}_0(\bkp ) - \left[ \vec{E}_0^*(\bkp ) \cdot \vec{Q}_0(\bkp ) \right] \vec{E}_0(\bkp )\right\}_3
  \nn\\
  &= \Re \frac{c^2}{8\pi\w} \int \!\dint^2\xp \; \alpha_1(\kp ) \left|\vec{E}_0(\bkp )\right|^2
  \nn\\
  &= S \frac{c^2}{8\pi\w} \alpha_1(\kp ) \left| \vec{E}_0(\bkp )\right|^2 .
  \label{eq:4.1}
\end{align}
In this result $S$ is the area of the $x_1x_2$ plane covered by the randomly rough surface, and the integrand in the first line is the time-averaged 3-component of the complex Poynting vector \cite{Simonsen2010}.
The minus sign on the right-hand side of the first equation compensates for the fact that the 3-component of the incident flux is negative, and we have used the fact that $\alpha_1(\kp )$ is real, so that $\vec{Q}_0(\bkp )$ is real, and $\vec{E}^*_0(\bkp ) \cdot \vec{Q}_0 (\bkp ) = 0$.

In a similar fashion we note that the total time-averaged transmitted flux is given by
\begin{align}
  P_\textrm{trans} =& -\Re\, \frac{c}{8\pi} \int \!\dint^2\xp\! \!\int\frac{\dint^2\qp}{(2\pi )^2}\! \int\frac{\dint^2\qp '}{(2\pi )^2} 
  \left\{ \vec{B}^*(\bqp ) \times \left[\frac{c}{\w} \vec{Q}_2(\bqp ') \times \vec{B}(\bqp ') \right]\right\}_3\nn\\
  & \qquad \qquad \qquad \times \exp \left\{ -\imu(\bqp - \bqp ')\cdot \pvec{x}  -\imu\left[\alpha_2(\qp ') - \alpha^*_2(\qp )\right]x_3    \right\}\nn\\
  %
  =& - \Re \frac{c^2}{8\pi\w} \int \!\frac{\dint^2\qp}{(2\pi )^2} \left\{ \vec{B}^*(\bqp ) \times \left[ \vec{Q}_2(\bqp )\times \vec{B}(\bqp ) \right] \right\}_3
    \, \exp \left[ 2 \Im \alpha_2(\qp ) x_3 \right] \nn\\
  =& - \Re \frac{c^2}{8\pi\w} \int \!\frac{\dint^2\qp}{(2\pi )^2}\,
        \left \{ \left| \vec{B}(\bqp )\right|^2 \vec{Q}_2 (\bqp )  - \left[ \vec{B}^*(\bqp )\cdot \vec{Q}_2(\bqp ) \right] \vec{B} (\bqp ) \right\}_3
       \exp [ 2 \Im \alpha_2(\qp ) x_3]\nn\\
 =&  \,\Re \frac{c^2}{32\pi^3\w} \int \dint^2\qp\;  \left|{\textbf B}(\bqp )\right|^2  \alpha_2(\qp)
        \exp [2 \Im\alpha_2(\qp ) x_3 ]
    \nn\\ &
   - \Re \frac{\imu c^4}{16\pi^2\e_2\w^3} \int \dint^2\qp\;  \Im\alpha_2(\qp )    
     \qp^2\left|B_p(\bqp )\right|^2 \exp [2\Im\alpha_2(\qp )x_3 ] .\label{eq:4.3}
\end{align}
\end{widetext}
The second term vanishes since it is the real part of a pure imaginary number. Thus we have
\begin{align}
  P_\textrm{trans} = \frac{c^2}{32\pi^3\w} \int\limits_{\qp < \sqrt{\e_2}\frac{\w}{c}} \dint^2\qp \; \alpha_2(\qp ) \left|\vec{B}(\bqp )\right|^2 .
  \label{eq:4.4}
\end{align}

The vectors $\bkp$ and $\bqp$ can be expressed in terms of the polar and azimuthal angles of incidence $(\theta_0,\phi_0)$ and transmission $(\theta_t,\phi_t)$, respectively, by
\begin{subequations}
  \label{eq:4.5}
  \begin{align}
    \bkp &= \sqrt{\e_1} \frac{\w}{c} \sin\theta_0 (\cos\phi_0,\sin\phi_0,0)
    \label{eq:4.5a}\\
    \bqp &= \sqrt{\e_2} \frac{\w}{c} \sin \theta_t (\cos\phi_t,\sin\phi_t, 0).
    \label{eq:4.5b}
  \end{align}
\end{subequations}
From these results it follows that
\begin{align}
  \dint^2\qp = \e_2 \left( \frac{\w}{c}\right)^2 \cos\theta_t \, \dint\Omega_t ,\label{eq:4.6}
\end{align}
where $\dint\Omega_t = \sin\theta_t \, \dint\theta_t \, \dint\phi_t$.  The total time-averaged transmitted flux becomes
\begin{align}
  P_\textrm{trans} &= \frac{\e^{3/2}_2\w^2}{32\pi^3c} \int \!\dint\Omega_t\; \cos^2\theta_t \left[ \left|B_p(\bqp )\right|^2 + \left|B_s(\bqp )\right|^2 \right].
\label{eq:4.7}
\end{align}
Similarly, the total time averaged incident flux, Eq.~(\ref{eq:4.1}), becomes
\begin{align}
  P_\textrm{inc} &= S\frac{\sqrt{\e_1}c}{8\pi} \cos\theta_0 \left[ \left|E_{0p}(\bkp )\right|^2 + \left|E_{0s}(\bkp )\right|^2\right] .\label{eq:4.8}
\end{align}
Thus by definition, the differential transmission coefficient is given by
\begin{align}
  \frac{\p T}{\p \Omega_t} &= \frac{1}{S} \frac{\e^{3/2}_{2}}{\e^{1/2}_{1}}  \left(\frac{\w}{2\pi c}\right)^2 \frac{\cos^2\theta_t}{\cos\theta_0} 
  \frac{\left|B_p(\bqp )\right|^2 + \left|B_s(\bqp )\right|^2}{ \left|E_{0p}(\bkp )\right|^2 + \left|E_{0s}(\bkp )\right|^2}.
\label{eq:4.9}
\end{align}
When we combine this result with Eq.~(\ref{eq:3.23}) we find that the contribution to the differential transmission coefficient when an incident plane wave of polarization  $\beta$, the projection of whose wave vector on the mean scattering plane is $\bkp$, is transmitted into a plane wave of polarization $\alpha$, the projection of whose wave vector on the mean scattering plane is $\bqp$, is given by
\begin{align}
  \frac{\p T_{\alpha\beta}(\bqp |\bkp )}{\p\Omega_t}
  &=
  \frac{1}{S} \frac{\e^{3/2}_2}{\e^{1/2}_1}\left( \frac{\w}{2\pi c}\right)^2 \frac{\cos^2\theta_t}{\cos\theta_0}
  \left| T_{\alpha\beta}(\bqp |\bkp )\right|^2 .
  \label{eq:4.10}
\end{align}

Since we are considering the transmission of light through a randomly rough interface, it is the average of this function over an ensemble of realizations of the surface profile function that we need to calculate.
This is the mean differential transmission coefficient, which is defined by
\begin{align}
  \left\la \frac{\p T_{\alpha\beta}(\bqp |\bkp )}{\p \Omega_t}\right\ra
  &=
  \frac{1}{S} \frac{\e^{3/2}_2}{\e^{1/2}_1} \left( \frac{\w}{2\pi c}\right)^2 \frac{\cos^2\theta_t}{\cos\theta_0}
  \left\la \left| T_{\alpha\beta} (\bqp |\bkp ) \right|^2\right\ra .
  \label{eq:4.11}
\end{align}
If we write the transmission amplitude $T_{\alpha\beta}(\bqp |\bkp )$ as the sum of its mean value and the fluctuation from this mean,
\begin{align}
  T_{\alpha\beta}(\bqp |\bkp )
  &=
  \left\la T_{\alpha\beta}(\bqp |\bkp )\right\ra  + \left[T_{\alpha\beta}(\bqp |\bkp ) - \left\la T_{\alpha\beta}(\bqp |\bkp )\right\ra \right] ,
  \label{eq:4.12}
\end{align}
then each of these two terms contributes separately to the mean differential transmission coefficient,
\begin{align}
  \left\la \frac{\p T_{\alpha\beta}(\bqp |\bkp )}{\p\Omega_t}\right\ra
  &=
  \left\la \frac{\p T_{\alpha\beta}(\bqp |\bkp )}{\p\Omega_t}\right\ra_\textrm{coh}
  + \left\la \frac{\p T_{\alpha\beta}(\bqp |\bkp )}{\p\Omega_t}\right\ra_\textrm{incoh} ,
  \label{eq:4.12-mislabeled}
\end{align}
where
\begin{align}
  \left\la \frac{\p T_{\alpha\beta}(\bqp |\bkp )}{\p \Omega_t}\right\ra_\textrm{coh}
  &=
  \frac{1}{S} \frac{\e^{3/2}_2}{\e^{1/2}_1} \left( \frac{\w}{2\pi c}\right)^2\frac{\cos^2\theta_t}{\cos\theta_0}
  \left|\la T_{\alpha\beta} (\bqp |\bkp )\ra \right|^2
  \label{eq:4.13}
\end{align}
and
\begin{align}
  &\left\la \frac{\p T_{\alpha\beta}(\bqp |\bkp )}{\p \Omega_t}\right\ra_\textrm{incoh}
  \nn\\
  &=
  \frac{1}{S} \frac{\e^{3/2}_2}{\e^{1/2}_1} \left( \frac{\w}{2\pi c}\right)^2\frac{\cos^2\theta_t}{\cos\theta_0} 
  \left[ \left\la  \left| T_{\alpha\beta} (\bqp |\bkp )-\la T_{\alpha\beta}(\bqp |\bkp )\ra \right|^2\right\ra \right] \nn\\
  &= \frac{1}{S} \frac{\e^{3/2}_2}{\e^{1/2}_1} \left( \frac{\w}{2\pi c}\right)^2 \frac{\cos^2\theta_t}{\cos\theta_0} 
  \left[ \left\la  \left| T_{\alpha\beta} (\bqp |\bkp )\right|^2\right\ra -\left|\Big\la T_{\alpha\beta}(\bqp |\bkp )\Big\ra \right|^2 \right]. \quad \label{eq:4.14}
\end{align}
The first contribution describes the refraction of the incident field, while the second contribution describes the diffuse transmission.

\section{Transmissivity and transmittance}
In the following we will refer to \textit{transmittance} as the fraction of the power flux incident on the rough surface that is transmitted through it, and \textit{transmissivity} as the fraction of the power flux incident on the rough surface that is transmitted coherently and co-polarized through it.
To obtain the transmissivity of the two-dimensional randomly rough interface we start with the result that
\begin{align}
  \la T_{\alpha\beta}(\bqp |\bkp )\ra = (2\pi )^2 \delta (\bqp - \bkp )\delta_{\alpha\beta}T_{\alpha}(\kp ) .\label{eq:4.15}
\end{align}
The presence of the delta function is due to the stationarity of the randomly rough surface; the Kronecker symbol $\delta_{\alpha\beta}$ arises from the conservation of angular momentum in the transmission process; and the result that $T_{\alpha}(\kp )$ depends on $\bkp$ only through its magnitude is due to the isotropy of the random roughness.

With the result given by Eq.~(\ref{eq:4.15}), the expression for $\la \p T_{\alpha\beta}(\bqp |\bkp )/\p\Omega_t \ra_\textrm{coh}$ given by Eq.~(\ref{eq:4.13}), becomes
\begin{align}
  \left\la \frac{\p T_{\alpha\alpha} (\pvec{q} |\pvec{k} )}{\p\Omega_t}\right\ra_\textrm{coh}
  &=
  \frac{\e^{3/2}_2}{\e^{1/2}_1} \left( \frac{\w}{c}\right)^2\frac{\cos^2\theta_t}{\cos\theta_0}
  \left| T_{\alpha}(\kp ) \right|^2 \, \delta (\pvec{q} - \pvec{k}),
  \label{eq:4.16}
\end{align}
where we have used the result
\begin{align}
  [(2\pi )^2\delta (\bqp - \bkp) ]^2
  &=
  (2\pi )^2 \delta (\vec{0}) \,
  (2\pi )^2\delta (\pvec{q} - \pvec{k}  )
  \nn\\
  &=
  S (2\pi )^2 \delta (\bqp - \bkp ) \label{eq:4.17}
\end{align}
in obtaining this expression.  We next use the result
\begin{align}
  \delta (\bqp - \bkp )
  &=
  \frac{1}{\kp } \delta ( \qp  - \kp )\, \delta (\phi_t - \phi_0)
  \nn \\
  &= \frac{1}{\sqrt{\e_1\e_2}} \left( \frac{c}{\omega} \right)^2
      \frac{ \delta ( \theta_t - \Theta_t )\, \delta (\phi_t - \phi_0) }{ \sin\theta_0 \cos\Theta_t  }
\label{eq:4.18-new}
\end{align}
to obtain
\begin{align}
  &\left\la \frac{\p T_{\alpha\alpha} (\bqp |\bkp )}{\p\Omega_t}\right\ra_\textrm{coh}
  \nn\\
  &\quad =
  \frac{\e_2}{\e_1} \frac{ \cos\Theta_t }{ \sin\theta_0 \cos\theta_0}
  \left| T_\alpha ( \kp ) \right|^2
  \delta ( \theta_t - \Theta_t )\, \delta (\phi_t - \phi_0 ),
  \label{eq:4.20-new}
\end{align}
where the the polar angle for the specular direction of transmission has, according to Snell's law, been denoted
\begin{align}
  \Theta_t \equiv \sin^{-1}\left(  \sqrt{ \frac{\e_1}{\e_2} } \sin\theta_0\right).
  \label{eq:4.19-new}
\end{align}
The transmissivity, ${\mathcal T}_{\alpha}(\theta_0)$, for light of $\alpha$ polarization is defined by
\begin{align}
  {\mathcal T}_{\alpha}(\theta_0)
  =& \int^{\frac{\pi}{2}}_{0} \dint\theta_t\, \sin\theta_t \int^{\pi}_{-\pi} \dint\phi_t\,
  \left\la \frac{T_{\alpha\alpha}(\bqp |\bkp )}{\p\Omega_t}\right\ra_\textrm{coh}
  \nn\\
  =& \frac{\e_2}{\e_1} \frac{ \cos\Theta_t \sin\Theta_t }{ \sin\theta_0 \cos\theta_0 }   \left| T_{\alpha}(\kp ) \right|^2
      \int_{0}^{\frac{\pi}{2}} \dint\theta_t \, \delta(\theta_t-\Theta_t)
  \nn\\
  =&
  \begin{cases}
    \displaystyle \sqrt{ \frac{\e_2}{\e_1} } \frac{\cos\Theta_t}{\cos\theta_0} \left| T_{\alpha}(\kp ) \right|^2,
    & 0 < \sqrt{\e_1/\e_2}\sin\theta_0 < 1
    \\
    0,
    & \mbox{otherwise}
  \end{cases}.
  \label{eq:4.21-new}
\end{align}
In writing this expression we have used the result that $\sin\Theta_t=\sqrt{\e_1/\e_2}\sin\theta_0$, and that  $\sin\theta_0$ is a monotonically increasing function of $\theta_0$ for $\ang{0}<\theta_0<\ang{90}$, and so therefore is $\sin\Theta_t$.
We see from Eq.~\eqref{eq:4.21-new} that when $\e_1>\e_2$ the transmissivity is nonzero for angles of incidence satisfying $0 < \theta_0 < \sin^{-1}({\sqrt{\e_2/\e_1}})$, and vanishes for angles of incidence satisfying $\sin^{-1}({\sqrt{\e_2/\e_1}}) < \theta_0 < \pi/2$. This result is a consequence for transmission of the existence of a critical angle for total internal reflection, namely
$\theta_0^\star=\sin^{-1}({\sqrt{\e_2/\e_1}})$.
In the case where $\e_1<\e_2$, the transmissivity is nonzero in the entire range of angles of incidence, $0<\theta_0<\pi/2$.

The function $T_{\alpha}(\kp )$ is obtained from Eq.~\eqref{eq:4.15}, with the aid of the result that $(2\pi)^2\delta(\vec{0})=S$, in the form
\begin{align}
  \label{eq:4.22}
  T_{\alpha}(\kp ) =   T_{\alpha}\left( \sqrt{ \e_1 }\frac{\w}{c}\sin\theta_0 \right)
                  = \frac{1}{S} \left< T_{\alpha\alpha}(\pvec{k} | \pvec{k}) \right>.
\end{align}

In addition to the transmissivity~\eqref{eq:4.21-new} that depends only on the co-polarized light transmitted coherently by the rough interface, it is also of interest to introduce the transmittance for light of $\beta$ polarization defined as
\begin{subequations}
  \label{eq:transmittance}
\begin{align}
  \label{eq:transmittance-A}
  {\mathscr T}_{\beta}(\theta_0)
  =&
  \sum_{\alpha=\mathrm{p},\mathrm{s}}  {\mathscr T}_{\alpha\beta}(\theta_0),
\end{align}
where
\begin{align}
  \label{eq:transmittance-B}
  {\mathscr T}_{\alpha\beta}(\theta_0)
  &=
  \int^{\frac{\pi}{2}}_{0} \dint\theta_t\, \sin\theta_t \int^{\pi}_{-\pi} \dint\phi_t\,
  \left< \frac{T_{\alpha\beta}(\bqp |\bkp )}{\p\Omega_t} \right>.
\end{align}
\end{subequations}
In light of Eq.~\eqref{eq:4.12-mislabeled}, the transmittance obtains contributions from light that has been transmitted coherently as well as  incoherently through the rough interface, ${\mathscr T}_{\beta}(\theta_0) = {\mathscr T}_{\beta}(\theta_0)_{\mathrm{coh}} + {\mathscr T}_{\beta}(\theta_0)_{\mathrm{incoh}}$, and both co- and cross-polarized transmitted light contribute to it.
Moreover, with Eq.~\eqref{eq:4.21-new}, and since cross-polarized coherently transmitted light is not allowed [see Eq.~\eqref{eq:4.15}], the coherent contribution to transmittance  for light of $\beta$ polarization equals the transmissivity for light of $\beta$ polarization; ${\mathscr T}_{\beta}(\theta_0)_{\mathrm{coh}} = {\mathcal T}_{\beta}(\theta_0)$. Therefore, Eq.~\eqref{eq:transmittance-A} can be written in the form
\begin{align}
  {\mathscr T}_{\beta}(\theta_0)  &=    {\mathcal T}_{\beta}(\theta_0)
    +  \sum_{\alpha=\mathrm{p},\mathrm{s}} {\mathscr T}_{\alpha\beta}(\theta_0)_{\mathrm{incoh}}.
    \label{eq:transmittance_sum}
\end{align}

It remains to remark that in cases where the incident light is not purely p- or s-polarized, the transmittance and transmissivity of the optical system will have to be calculated on the basis of weighted sums of the expressions in Eqs.~\eqref{eq:4.21-new} and  \eqref{eq:transmittance} where the weights reflect the fraction of p and s polarization associated with the incident light.

\section{Results and discussions}

\begin{figure*}
  \centering
  \includegraphics[width=0.47\textwidth]{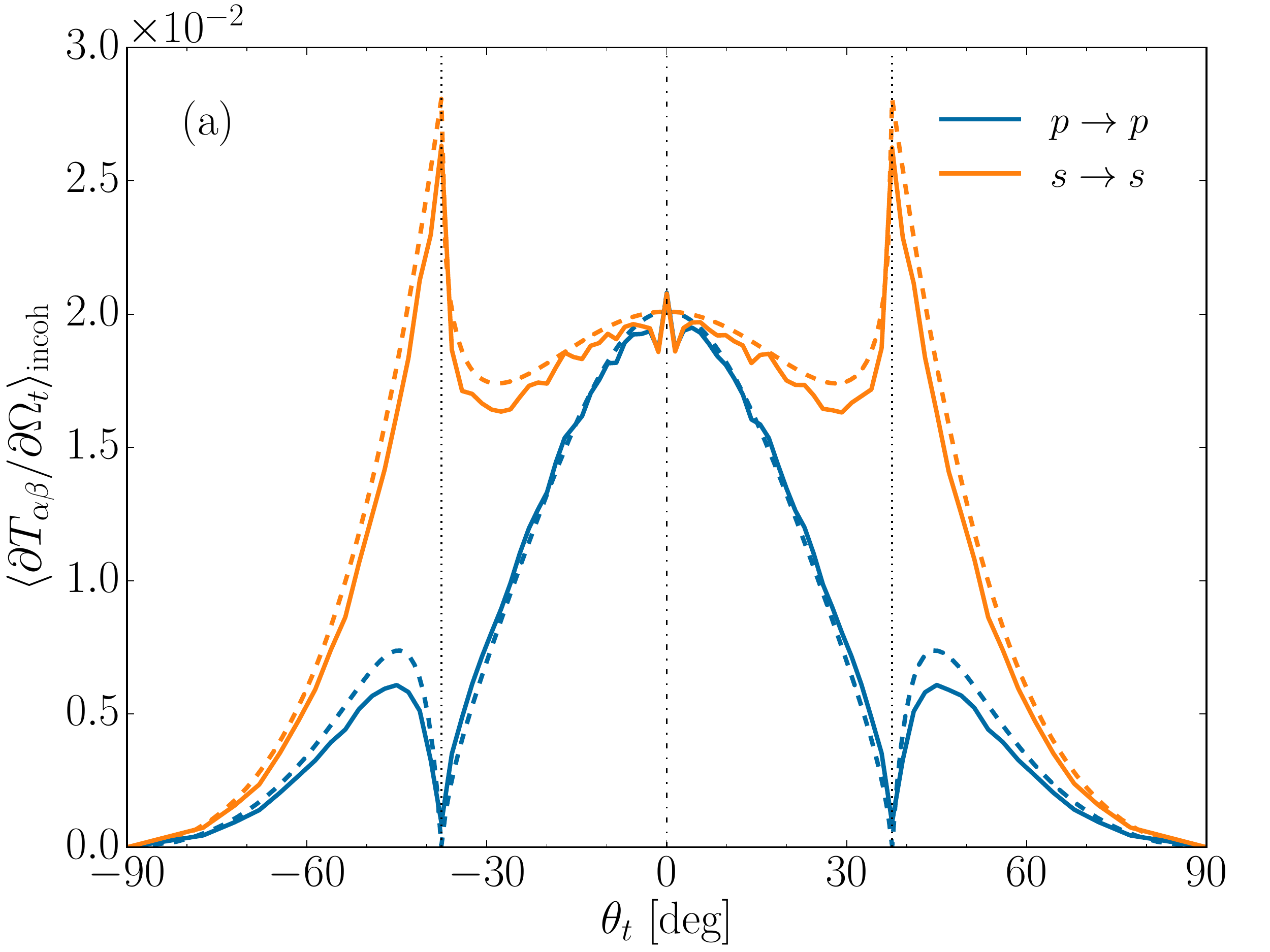}
  \includegraphics[width=0.47\textwidth]{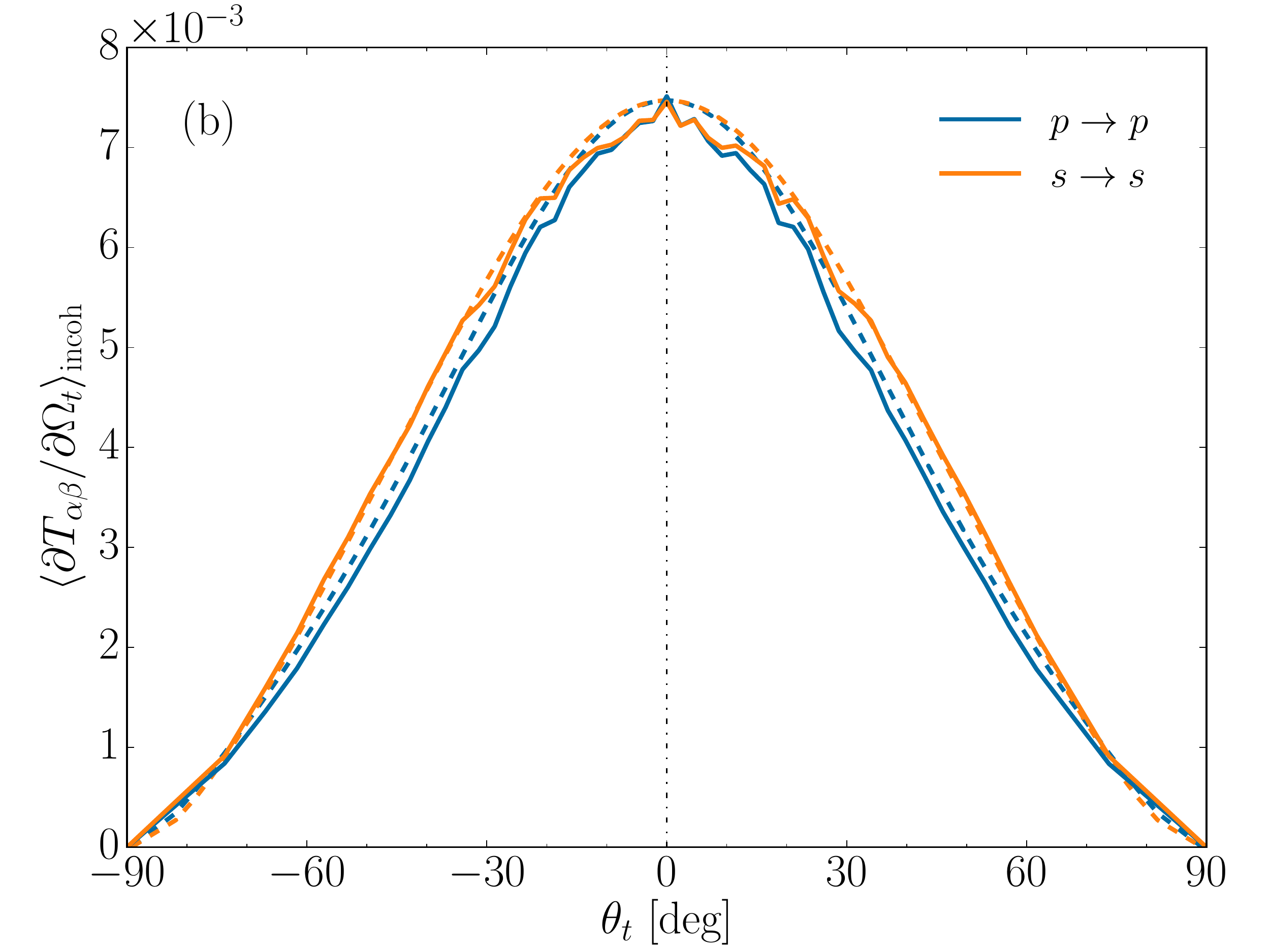}
   \caption{The contribution to the incoherent component of the mean differential transmission coefficient from the in-plane, co-polarized transmission of p- and s-polarized light incident normally [$(\theta_0,\phi_0)=(\ang{0},\ang{0})$] on the random vacuum-dielectric interface, as a function of the angle of transmission $\theta_t$. (a) The medium of incidence is vacuum [$\e_1=1$; $\e_2=2.6896$]; (b) The medium of incidence is the dielectric [$\e_1=2.6896$; $\e_2=1$].
   Negative values of $\theta_t$ correspond to light transmitted in the azimuthal direction of $\phi_t=\ang{180}$.
   Results for (in-plane) cross-polarized transmission have not been indicated since they are generally suppressed in the plane-of-incidence. The results presented as solid lines were obtained on the basis of numerical solutions of the reduced Rayleigh equation~\eqref{eq:3.24} for an ensemble of \num{5000} surface realizations. The dashed curves represent the result of the small amplitude perturbation theory~\eqref{eq:5.20} to first order, assuming polarization as indicated for the solid lines of the same color.
   The specular direction of transmission is indicated by the vertical dash-dotted line at $\theta_t=\ang{0}$, and in Fig.~\protect\ref{fig:inplane_mdtc_theta_0}(a), the vertical dotted lines at $\theta_t=\pm\theta_t^\star$ indicate the position of the critical angle where $\theta_t^\star=\sin^{-1}(\sqrt{\e_1/\e_2})\approx\ang{37.6}$ for the parameters assumed. The wavelength of the incident light in vacuum was $\lambda$.
   The rough interface was assumed to have a root-mean-square roughness of $\delta = \lambda/20$, and it was characterized by  an isotropic Gaussian power spectrum~\eqref{eq:2.3} of transverse correlation length $a=\lambda/4$. In the numerical calculations it was assumed that the surface covered an area $L\times L$, with $L=25 \lambda$, and the surface was discretized on a grid of $321 \times 321$ points.
}
\label{fig:inplane_mdtc_theta_0}
\end{figure*}

Calculations were carried out for two-dimensional randomly rough dielectric surfaces defined by an isotropic Gaussian height distribution of rms height $\delta=\lambda/20$ and an isotropic Gaussian correlation function of transverse correlation length $a=\lambda/4$. The incident light consisted of a p- or s-polarized plane wave of wavelength $\lambda$ (in vacuum) and well-defined angles of incidence $(\theta_0,\phi_0)$.
The dielectric medium was assumed to be a photoresist defined by the dielectric constant $\varepsilon=2.6896$.
The azimuthal angle of incidence was $\phi_0=0^{\circ}$ in all simulation results presented in this work; this choice for $\phi_0$ is somewhat arbitrary, since, due to the isotropy of the roughness, results for another choice of $\phi_0$ can be obtained from the results presented here by a trivial rotation.
Realizations of the  surface profile function $\zxp$ were generated~\cite{Maradudin1990,Simonsen2011} on a grid of $N_x\times N_x =321\times321$ points.
The surfaces covered a square region of the $x_1 x_2$ plane of edge $L=25\lambda$, giving an area  $S=L^2$.
With these spatial parameters, the corresponding momentum space parameters used in the simulations were $\Delta q=2\pi/L$~for the discretization intervals in momentum space, and the largest momentum value that was resolved was $\mathcal{Q}=6.4\omega/c$.

The reduced Rayleigh equation~\eqref{eq:3.24} was solved numerically by the method described in detail in Ref.~\citenum{Nordam2013a}, so only a summary of this method will be presented here.
In evaluating the $\pvec{q}$ integral in Eq.~\eqref{eq:3.24}, the infinite limits of integration were replaced by finite limits $|\pvec{q}| < {\mathcal Q}/2$, and the integration was carried out by a two-dimensional version of the extended midpoint rule \cite[p.~161]{Book:Press1996} applied to the circular subsection of a grid in the $q_1q_2$ plane which is determined by the Nyquist sampling theorem \cite[p.~605]{Book:Press1996} and the properties of the  discrete Fourier transform~\cite{Nordam2013a}.
The function $I(\gamma |\pvec{q} )$ was evaluated by expanding the integrand in Eq.~\eqref{eq:3.12-is} in powers of $\zeta (\pvec{x} )$ and calculating the Fourier transform of $\zeta^n(\pvec{x} )$ by the fast Fourier transform~\cite{Nordam2013a}.
For these expansions we used the first $\mathcal{N}=18$ terms.
The resulting matrix equations were solved by LU factorization and back substitution, using the ScaLAPACK library~\cite{scalapack}.

These calculations  were carried out for a large number $N_p$ of  realizations of the surface profile function $\zxp$ for an incident plane wave of p or s polarization.  For each surface realization the transmission amplitude $T_{\alpha\beta}(\pvec{q} |\pvec{k} )$ and its squared modulus $|T_{\alpha\beta}(\pvec{q} |\pvec{k} )|^2$ were obtained.
An arithmetic average of the $N_p$ results for these quantities yielded the mean values $\la T_{\alpha\beta}(\bqp \bkp )\ra $ and $\la |T_{\alpha\beta}(\bqp |\bkp )|^2\ra$ entering Eq.~\eqref{eq:4.14} for the mean differential transmission coefficient, and related quantities [see Eqs.~\eqref{eq:4.22} and \eqref{eq:transmittance_sum}].

Investigating the energy conservation of our simulation results can be a useful test of their accuracy. In combining simulation results from the current work with corresponding results obtained for the mean differential reflection coefficient $\left< \partial R_{\alpha\beta}/\partial \Omega_s\right>$ through the use of the computationally similar methods presented in Ref.~\citenum{Hetland2016a}, we may add the total reflected and transmitted power for any lossless system.
When the reflectance is added to the transmittance for any of the systems investigated in the current work, it is found that the results of these calculations satisfy unitarity with an error smaller than $10^{-4}$. This testifies to the accuracy of the approach used, and it is also a good indicator for satisfactory discretization. It should be noted, however, that unitarity is a necessary, but not sufficient, condition for the correctness of the presented results.
Through a preliminary investigation, unitarity seemed to be satisfied to a satisfactory degree for surfaces with a root mean square roughness up to about two times larger than the roughness used in obtaining the results presented in this paper, if the correlation function was kept the same.

\subsection{Normal incidence}
In Fig.~\ref{fig:inplane_mdtc_theta_0} we display the mean differential transmission coefficient (MDTC) in the plane of incidence as a function of the polar angle of transmission when the random surface is illuminated from the vacuum at normal incidence by p- and s-polarized light, Fig.~\ref{fig:inplane_mdtc_theta_0}(a), and when it is illuminated from the  dielectric medium, Fig.~\ref{fig:inplane_mdtc_theta_0}(b).
Only results for in-plane~[$\pvec{q}\parallel\pvec{k}$] co-polarized transmission are presented, since in-plane cross-polarized transmission is suppressed due to the absence of a contribution from single-scattering processes. An ensemble of \num{5000} realizations of the surface profile function was used to produce the averaged results presented in each of these figures.

From Fig.~\ref{fig:inplane_mdtc_theta_0}(a) it is observed that the curves display both maxima and minima in the $\ppol\to\ppol$ transmission spectrum, and peaks in the $\spol\to\spol$ transmission spectrum. In contrast, the curves presented in Fig.~\ref{fig:inplane_mdtc_theta_0}(b) are featureless, and are nearly identical.
The presence of these features, and others in subsequent figures, can be understood if we calculate the contribution to the MDTC from the light transmitted incoherently through the random interface as an expansion in powers of the surface profile function. This calculation, outlined in Appendix~\ref{app:SAPT}, yields the result that to lowest nonzero order in $\zeta(\pvec{x})$ we have
\begin{widetext}
\begin{subequations}
\label{eq:5.20}
\begin{align}
\left\la \frac{\p T_{pp}(\pvec{q} | \pvec{k} )}{\p\Omega_t}\right\ra_\textrm{incoh}
   &=
      \frac{\delta^2}{\pi^2} (\e_2-\e_1)^2 \e^{1/2}_1\e^{5/2}_2   
      \left( \frac{\w}{c} \right)^2 \frac{\cos^2\theta_t}{\cos\theta_0} g(|\bqp -\bkp |)
      \frac{1}{|d_p(\qp )|^2} \left|\alpha_1(\qp ) (\pvecUnit{q}\cdot \pvecUnit{k})  \alpha_2(\kp ) + \qp \kp \right|^2 \frac{\alpha^2_1(\kp )}{|d_p(\kp )|^2}
     \label{eq:5.20A}
     \\
\left\la \frac{\p T_{ps}(\pvec{q} | \pvec{k} )}{\p\Omega_t}\right\ra_\textrm{incoh}
   &=
      \frac{\delta^2}{\pi^2} (\e_2-\e_1)^2 \frac{\e^{5/2}_2}{\e^{1/2}_1}   
      \left( \frac{\w}{c} \right)^4 \frac{\cos^2\theta_t}{\cos\theta_0} g(|\bqp -\bkp |)
      \frac{ \left| \alpha_1(\qp ) \right|^2  }{|d_p(\qp )|^2}
      \left( \left[ \pvecUnit{q} \times \pvecUnit{k} \right]_3\right)^2
       \frac{  \alpha^2_1(\kp ) }{|d_s(\kp )|^2}
     \label{eq:5.20B}
     \\
\left\la \frac{\p T_{sp}(\pvec{q} | \pvec{k} )}{\p\Omega_t}\right\ra_\textrm{incoh}
   &=
      \frac{\delta^2}{\pi^2} (\e_2-\e_1)^2 \frac{\e^{1/2}_2}{\e^{1/2}_1}   
      \left( \frac{\w}{c} \right)^4 \frac{\cos^2\theta_t}{\cos\theta_0} g(|\bqp -\bkp |)
      \frac{ 1  }{|d_s(\qp )|^2}
      \left( \left[ \pvecUnit{q} \times \pvecUnit{k} \right]_3\right)^2
       \frac{ \alpha^2_1(\kp ) \left| \alpha_2(\kp ) \right|^2 }{|d_p(\kp )|^2}
     \label{eq:5.20C}
  \\
 \left\la \frac{\p T_{ss}(\bqp |\bkp )}{\p\Omega_t}\right\ra_\textrm{incoh}
   &=
    \frac{\delta^2}{\pi^2} (\e_2-\e_1)^2 \frac{\e^{3/2}_2}{\e^{1/2}_1} 
    \left( \frac{\w}{c}\right)^6 \frac{\cos^2\theta_t}{\cos\theta_0} g(|\bqp - \bkp |)
      \frac{1}{|d_s(\qp )|^2} (\pvecUnit{q}\cdot \pvecUnit{k})^2  \frac{\alpha^2_1(\kp )}{|d_s(\kp )|^2},
    \label{eq:5.20D}
\end{align}
\end{subequations}
\end{widetext}
where the functions $d_\alpha(\qp)$ and $d_\alpha(\kp)$ for $\alpha=p,s$ are presented in Eqs.~\eqref{app:eq:A11} and \eqref{eq:5.21}. In the following we will refer to Eq.~\eqref{eq:5.20} as the results of \textit{small amplitude perturbation theory} (SAPT) to first order.
Results from numerical evaluations of Eq.~\eqref{eq:5.20} for normal incidence and in-plane transmission [$\pvecUnit{q}\parallel \pvecUnit{k}$] are displayed as dashed lines in Fig.~\ref{fig:inplane_mdtc_theta_0} and several figures to follow.
For Fig.~\ref{fig:inplane_mdtc_theta_0} we have not included results for transmission \textit{out-of-plane} [$\pvecUnit{q}\cdot\pvecUnit{k}=0$], since, for normal incidence, the results for co-polarized in-plane transmission are identical with the results for cross-polarized out-of-plane transmission.
We notice in passing that the unit vectors $\pvecUnit{q}=\pvec{q}/\qp$ and $\pvecUnit{k}=\pvec{k}/\kp$ are well defined also for  $\theta_t=\ang{0}$ and $\theta_0=\ang{0}$, respectively, as follows from Eq.~\eqref{eq:4.5}.

From Fig.~\ref{fig:inplane_mdtc_theta_0} it is observed that the single-scattering perturbation theory reproduces fairly well the overall shape of the MDTC for in-plane co-polarized transmission, at least for the level of roughness assumed in producing these results. However, there is a difference in amplitude between the simulation results and the curves produced from perturbation theory, in particular when $\e_1<\e_2$.

The results from SAPT can be further analyzed in order to understand all features seen in Fig.~\ref{fig:inplane_mdtc_theta_0}. With the aid of $\qp=\sqrt{\e_2}(\omega/c)\sin\theta_t$, $d_\alpha(\qp)$ can be written in the form
\begin{subequations}
\label{eq:5.21}
\begin{align}
d_p(\qp ) 
          &=\sqrt{\e_2}\frac{\w}{c} \bigg\{ \e_2 \bigg[\bigg( \frac{\e_1-\e_2}{\e_2}\bigg) + \cos^2\theta_t\bigg]\sfr + \e_1 \cos\theta_t \bigg\}
 \label{eq:5.21a}
 \\
d_s(\qp ) 
          &= \sqrt{\e_2} \frac{\w}{c} \bigg\{ \bigg[ \bigg( \frac{\e_1-\e_2}{\e_2}\bigg) + \cos^2\theta_t\bigg]\sfr + \cos\theta_t\bigg\},
\label{eq:5.21b}
 %
 %
\end{align}
and from $\kp=\sqrt{\e_1}(\omega/c)\sin\theta_0$, $d_\alpha(\kp)$ can be expressed as
\begin{align}
d_p(\kp ) 
          &=\sqrt{\e_1}\frac{\w}{c} \bigg\{ \e_1 \bigg[\bigg( \frac{\e_2-\e_1}{\e_1}\bigg) + \cos^2\theta_0\bigg]\sfr + \e_2 \cos\theta_0 \bigg\}
 \label{eq:5.21c}
 \\
d_s(\kp ) 
          &= \sqrt{\e_1} \frac{\w}{c} \bigg\{ \bigg[ \bigg( \frac{\e_2-\e_1}{\e_1}\bigg) + \cos^2\theta_0\bigg]\sfr + \cos\theta_0\bigg\}.
\label{eq:5.21d}
\end{align}
\end{subequations}
We see from Eqs.~\eqref{eq:5.21a}  and ~\eqref{eq:5.21b} that when $\e_1$ is greater than $\e_2$, both $d_p(\qp )$ and $d_s(\qp )$ are real continuous monotonically decreasing functions of $\theta_t$, and so therefore are $|d_p(\qp )|^2$ and $|d_s(\qp )|^2$.  This leads to smooth dependencies of the MDTC on the angle of transmission~[Fig.~\ref{fig:inplane_mdtc_theta_0}(b)].
However, when $\e_1$ is smaller than $\e_2$, the first term in the expressions for $d_p(\qp )$ and $d_s (\qp )$ vanishes for a polar angle of transmission $\theta_t=\theta_t^\star$ defined by $\cos\theta_t^\star = [(\e_2-\e_1)/\e_2]\sfr$, or, equivalently, when $\sin\theta_t^\star = \sqrt{\e_1/\e_2}$, and becomes pure imaginary as $\theta_t$ increases beyond the angle 
\begin{align}
  \theta_t^\star=\sin^{-1}\sqrt{ \frac{\e_1}{\e_2} },
\end{align}
which is the critical angle for total internal reflection in the corresponding, inverse, flat-surface system where $\e_1\to\e_2$ and $\e_2\to\e_1$.
The functions $|d_p(\qp )|^{-2}$ and $|d_s(\qp)|^{-2}$ in Eq.~\eqref{eq:5.20} therefore display asymmetric peaks at the polar angle of transmission $\theta_t=\theta_t^\star$.
For $\spol\to\spol$ co-polarized in-plane (incoherent) transmission at normal incidence we therefore see sharp peaks in the MDTC at this polar angle both for forward and backward scattered light~[Fig.~\ref{fig:inplane_mdtc_theta_0}(a)]. The same peaks will then also be visible for $\ppol\to\spol$ cross-polarized out-of-plane transmission at normal incidence.
However, in the case of $\ppol\to\ppol$ co-polarized transmission we instead see dips at $\theta_t^\star$ in Fig.~\ref{fig:inplane_mdtc_theta_0}(a). In the case of the first-order SAPT results, the MDTC does indeed go to zero at this ``critical'' polar angle. This is due to the zeros in Eq.~\eqref{eq:5.20A}, specifically the zeros in the function
\begin{align}
  F(\bqp|\bkp) = \left|\alpha_1(\qp ) (\pvecUnit{q}\cdot \pvecUnit{k})  \alpha_2(\kp ) + \qp \kp \right|^2.
  \label{eq:F}
\end{align}
For normal incidence [$\kp=0$] and in-plane transmission [$\pvec{q}\parallel\pvec{k}$], the function $F(\bqp|\bkp)$ is zero for $\a_1(\qp)=0$.
This is the case for $\qp=\sqrt{\e_1}\w/c$ [Eq.~\eqref{eq:3.4}], which corresponds to $\theta_t=\theta_t^\star$ in the medium of transmission when $\e_2$ is greater than $\e_1$.
Finally, in the case of $\spol\to\ppol$ cross-polarized transmission, we will also see dips at $\theta_t^\star$ due to the simple factor $\a_1(\qp)$ in Eq.~\eqref{eq:5.20B}, but this factor is zero at this angle of transmission regardless of the angle of incidence.

The peaks observed in Fig.~\ref{fig:inplane_mdtc_theta_0}(a) where $\e_1<\e_2$ are the optical analogues of the \textit{Yoneda peaks} observed in the scattering (in reflection) of x-rays from both
metallic~\cite{Yoneda1963,Sinha1988,Gorodnichev1988,Renaud2009} and non-metallic~\cite{Dosch1987,Stepanov2000,Kitahara2002,Gasse2016} surfaces, later described as ``quasi-anomalous scattering peaks'' in the two-dimensional numerical work by Kawanishi~\etal~\cite{Kawanishi1997}.
The Yoneda peaks were originally observed as sharp peaks for incidence close to the grazing angle, as the difference in the dielectric constants of the two scattering media is very small at x-ray frequencies.
In the following, by Yoneda peaks we will mean well-defined maxima in the angular distribution of the intensity of the transmitted light at, or slightly above, the critical polar angle in the medium of transmission for which the wavenumber turns non-propagating in the medium of incidence, when $\e_1 < \e_2$.
A more detailed discussion on Yoneda peaks in reflection and in general can be found in Ref.~\citenum{Hetland2016a}.

Because the Yoneda peaks and the minima given by Eq.~\eqref{eq:F} are present in the expressions for the MDTC obtained in the lowest order in the surface profile function, the second, they can be interpreted as single-scattering phenomena, not multiple-scattering effects. This is supported by the qualitative similarity between the plots presented in Fig.~\ref{fig:inplane_mdtc_theta_0}.
We specify that the polar angle of transmission where the Yoneda phenomenon can be observed is determined only by the ratio of the dielectric constants of the two media; it does not, for instance, depend on the polar angle of incidence.


\begin{figure*}
  \centering
  \includegraphics[width=0.8\textwidth]{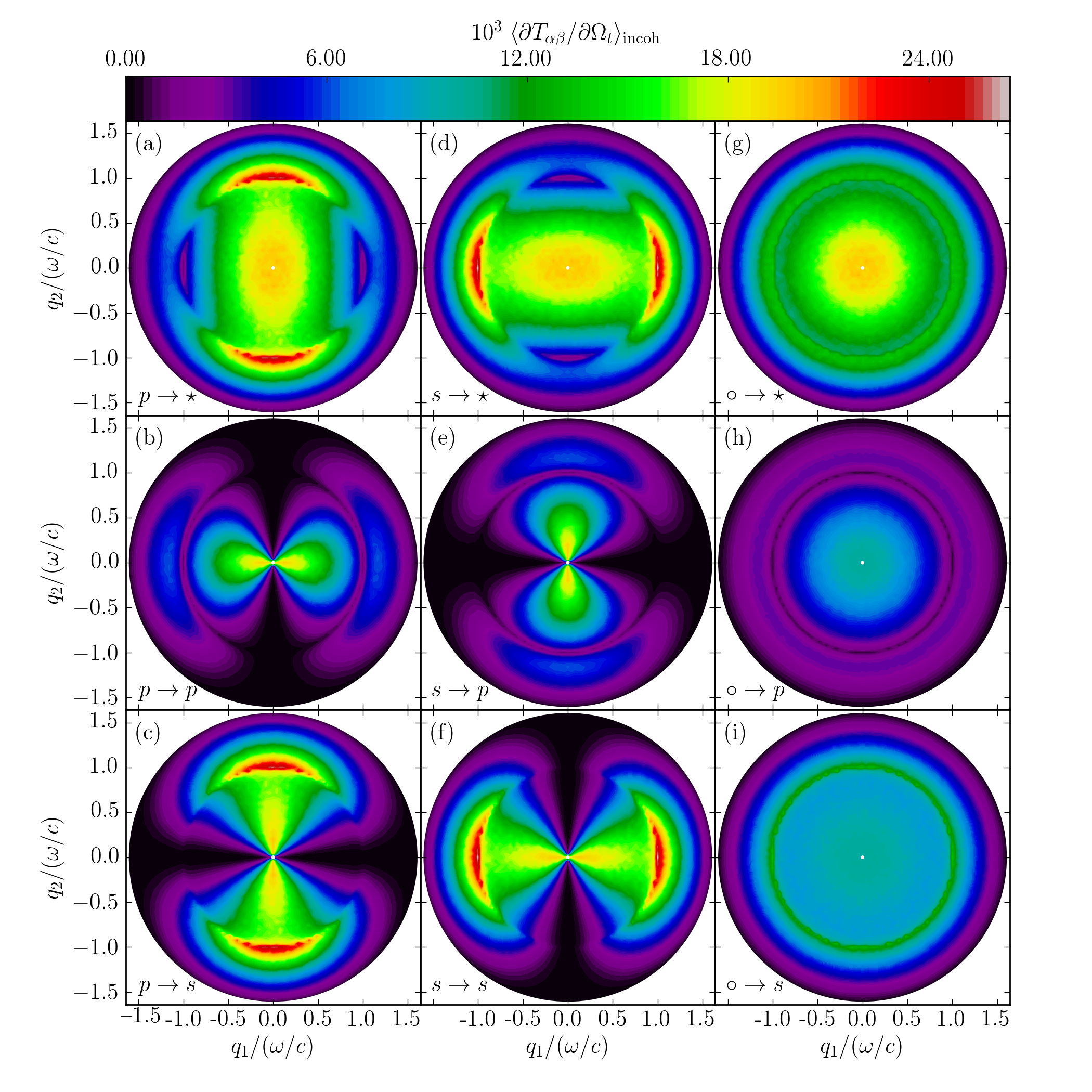}
  \caption{
  The incoherent component of the mean differential transmission coefficient, showing the full angular intensity distribution as a function of the lateral wave vector of the light transmitted from vacuum into a dielectric medium separated by a rough interface.
  The angles of incidence are $(\theta_0,\phi_0)=(\ang{0},\ang{0})$. Notice the rapid changes in intensity around the polar angle $\theta_t=\theta_t^\star=\sin^{-1}(\sqrt{\e_1/\e_2})$ corresponding to $\qp=\sqrt{\e_1}\omega/c$.
  The position of the specular direction in transmission is indicated by white dots. The parameters assumed for the scattering geometry and used in performing the numerical simulations have values that are identical to those assumed in obtaining the results of Fig.~\protect\ref{fig:inplane_mdtc_theta_0}(a). The in-plane intensity variations in Figs.~\protect\ref{fig:2Dmdtc_vtp_theta_0}(b) and ~\protect\ref{fig:2Dmdtc_vtp_theta_0}(f) are the curves depicted in Fig.~\protect\ref{fig:inplane_mdtc_theta_0}(a).
  The star notation, \textit{e.g.} $p\rightarrow\star$, indicates that the polarization of the transmitted light was not recorded. Furthermore, in \textit{e.g} Fig.~\protect\ref{fig:2Dmdtc_vtp_theta_0}(g), the open circle in $\circ\rightarrow\star$ symbolizes that the incident light was unpolarized; this simulation result was obtained by adding \textit{half} of the results from Figs.~\protect\ref{fig:2Dmdtc_vtp_theta_0}(a) and ~\protect\ref{fig:2Dmdtc_vtp_theta_0}(d).
  [Parameters: $\e_1=1.0$, $\e_2=2.6896$; $\delta=\lambda/20$, $a=\lambda/4$].
  }
\label{fig:2Dmdtc_vtp_theta_0}
\end{figure*}
%
\begin{figure*}
  \centering
  \includegraphics[width=0.8\textwidth]{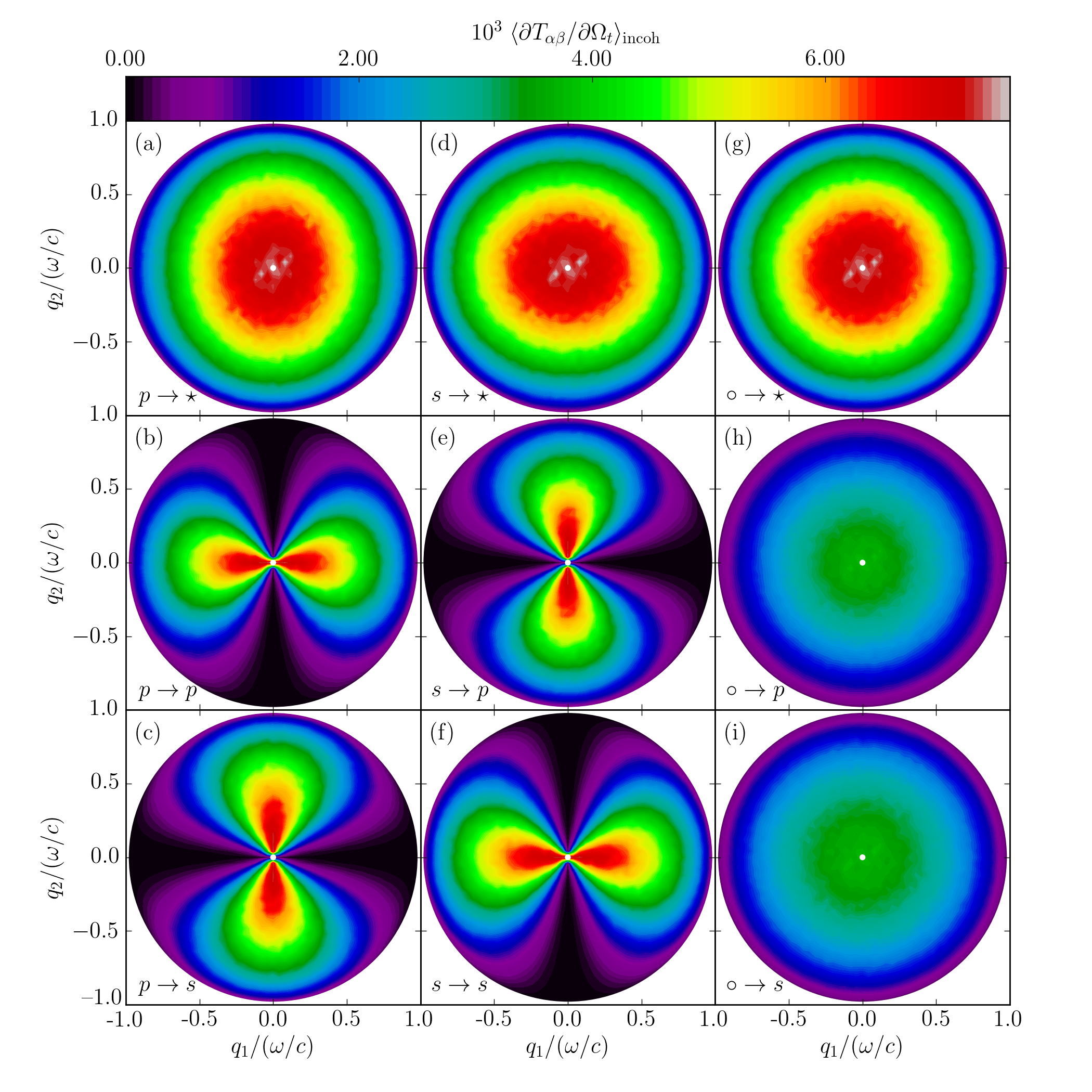}
  \caption{
  Same as Fig.~\protect\ref{fig:2Dmdtc_vtp_theta_0}, but for light incident from the dielectric side onto the interface with vacuum.
  The in-plane intensity variations in Figs.~\protect\ref{fig:2Dmdtc_ptv_theta_0}(b) and ~\protect\ref{fig:2Dmdtc_ptv_theta_0}(f) are the curves depicted in Fig.~\protect\ref{fig:inplane_mdtc_theta_0}(b).
  [Parameters: $\e_1=2.6896$, $\e_2=1.0$; $\delta=\lambda/20$, $a=\lambda/4$].
  }
 \label{fig:2Dmdtc_ptv_theta_0}
\end{figure*}

\smallskip
We now turn to the angular intensity distributions of the transmitted light. In Figs.~\ref{fig:2Dmdtc_vtp_theta_0} and \ref{fig:2Dmdtc_ptv_theta_0} we present simulation results for the contribution to the MDTC from the light that has been transmitted incoherently through the randomly rough interface, that display the full angular distribution of this contribution.
These two figures were obtained under the assumption that the angles of incidence were $(\theta_0,\phi_0)=(\ang{0},\ang{0})$; cuts along the plane of incidence of these angular intensity distributions result in the curves presented in Fig.~\ref{fig:inplane_mdtc_theta_0}.
Therefore, the parameters assumed in producing the results of Figs.~\ref{fig:inplane_mdtc_theta_0}(a)  and \ref{fig:2Dmdtc_vtp_theta_0} are identical, and so are the parameters assumed in obtaining Figs.~\ref{fig:inplane_mdtc_theta_0}(b)  and \ref{fig:2Dmdtc_ptv_theta_0}.

All angular intensity distributions presented in this work, including those in Figs.~\ref{fig:2Dmdtc_vtp_theta_0} and \ref{fig:2Dmdtc_ptv_theta_0}, are organized in the same fashion. They are arranged in $3\times 3$ subfigures where each row and column of the array correspond to the angular distribution of the incoherent component of the mean differential transmission coefficient for a given state of polarization of the transmitted and incident light, respectively.
The lower left $2\times 2$ corner of such figures corresponds to the cases where $\beta$-polarized incident light is transmitted by the rough interface into $\alpha$-polarized light, denoted $\beta\rightarrow\alpha$ in the lower left corner of each subfigure, where $\alpha=p,s$ and the same for $\beta$. Moreover, the first row corresponds to results where the polarization of the transmitted light was not recorded (indicated by $\star$); such results are obtained by adding the other two results from the same column.
The last column of the angular intensity distribution figures corresponds to the situation when the incident light is \textit{unpolarized} (indicated by an open circle, $\circ$); these results are obtained by adding \textit{half} of the other two results present in the same row. For instance, the subfigure in the upper right corner, labeled $\circ\rightarrow\star$, refers to unpolarized light (the open circle) transmitted by the surface into light for which we do not record the polarization (the star). It should be stressed that even if the polarization of the transmitted light is not recorded, it does not mean that the transmitted light is unpolarized; in general this is not the case as can be seen by, for instance, inspecting Fig.~\ref{fig:2Dmdtc_vtp_theta_0}.

When both the incident and transmitted light are linearly polarized, the lower left $2\times 2$ corners of Figs.~\ref{fig:2Dmdtc_vtp_theta_0} and \ref{fig:2Dmdtc_ptv_theta_0} show that the angular distributions of the incoherent component of the mean differential transmission coefficients take on dipole-like patterns oriented along the plane-of-incidence for co-polarization and perpendicular to it for cross-polarization. We note that such patterns are a consequence of our definitions of the polarization vectors, and that similar patterns have recently been observed in reflection~\cite{Nordam2013a,Nordam2014,Hetland2016a}.
It was already concluded based on Fig.~\ref{fig:inplane_mdtc_theta_0} that the in-plane, co-polarized transmission is rather different for  p and s polarization when the medium of incidence is vacuum, and rather similar when the medium of incidence is the dielectric. Not surprisingly, a similar conclusion can be drawn by inspecting the co-polarized angular intensity distributions depicted in the $\beta\to\beta$ subfigures of Figs.~\ref{fig:2Dmdtc_vtp_theta_0} and \ref{fig:2Dmdtc_ptv_theta_0} [$\beta=\ppol,\spol$].
For normal incidence, the angular intensity distributions for cross- and co-polarized transmission are intimately related to each other, but only if they share the same polarization state of the transmitted light; in fact, the former distributions are \ang{90} rotations of the latter.
For instance for scattering into s-polarized light, this can be understood if we note from Eqs.~\eqref{eq:5.20C}, \eqref{eq:5.20D} and \eqref{eq:3.25}~[see also Eq.~\eqref{app:eq:A13} of Appendix~\ref{app:SAPT}] that to the lowest nonzero order  in $\zeta(\pvec{x})$ we have
\begin{subequations}
\begin{align}
    \left\la \frac{\p T_{sp}(\pvec{q} | \pvec{k} )}{\p\Omega_t}\right\ra_\textrm{incoh}
   &=
      \frac{\delta^2}{\pi^2} (\e_2-\e_1)^2 \e^{1/2}_1 \e^{5/2}_2
      \left( \frac{\w}{c} \right)^2 \frac{\cos^2\theta_t}{\cos\theta_0}
      \nn\\
      &\quad\times g(|\bqp -\bkp |)
      \frac{ \left|M_{sp}(\pvec{q}|\pvec{k})\right|^2 \alpha^2_1(\kp ) }{ |d_s(\qp )|^2 |d_p(\kp )|^2},
     \label{eq:5.20C-M}
  \\
 \left\la \frac{\p T_{ss}(\bqp |\bkp )}{\p\Omega_t}\right\ra_\textrm{incoh}
   &=
    \frac{\delta^2}{\pi^2} (\e_2-\e_1)^2 \frac{\e^{3/2}_2}{\e^{1/2}_1} 
    \left( \frac{\w}{c}\right)^2 \frac{\cos^2\theta_t}{\cos\theta_0}
    \nn\\
    &\quad\times g(|\bqp - \bkp |)
      \frac{ \left|M_{ss}(\pvec{q}|\pvec{k})\right|^2 \alpha^2_1(\kp ) }{ |d_s(\qp )|^2 |d_s(\kp )|^2},
    \label{eq:5.20D-M}
\end{align}
\end{subequations}
where the matrix elements $M_{sp}(\pvec{q}|\pvec{k})$ and $M_{ss}(\pvec{q}|\pvec{k})$ are presented in Eq.~\eqref{eq:3.25}. For normal incidence, $d_p(0)/ \sqrt{\e_1\e_2}  = d_s(0)$ and $M_{\spol\ppol}(\pvec{q}|\vec{0})$ out-of-plane equals   $M_{\spol\spol}(\pvec{q}|\vec{0})$ in-plane.
This means that $\la \partial T_{\spol\ppol}(\pvec{q} |\vec{0}) / \partial\Omega_t  \ra_\textrm{incoh}$ will equal $\la \partial T_{\spol\spol}(\pvec{q}' |\vec{0}) / \partial\Omega_t \ra_\textrm{incoh}$ if $\pvec{q}$, after a rotation by an angle of \ang{90}, equals $\pvec{q}'$.
A similar argument can be used to relate the angular distribution of $\la \partial T_{\ppol\spol}(\pvec{q} |\vec{0}) / \partial\Omega_t  \ra_\textrm{incoh}$ to a \ang{90} rotation of the angular distribution of  $\la \partial T_{\ppol\ppol}(\pvec{q} |\vec{0}) / \partial\Omega_t  \ra_\textrm{incoh}$.
This symmetry property of the angular intensity distributions at normal incidence is readily observed in Figs.~\ref{fig:2Dmdtc_vtp_theta_0} and \ref{fig:2Dmdtc_ptv_theta_0}. Hence, we conclude that the regions of high intensity observed in the cross-polarized angular intensity distribution in Fig.~\ref{fig:2Dmdtc_vtp_theta_0}(c) around the out-of-plane direction are also Yoneda peaks; their origin is due to the peaking factor $|d_s(\qp)|^{-2}$ vs. transmitted wave-number, identical to what we found for the in-plane peaks in the co-polarized transmitted light.

When $\e_1<\e_2$, Yoneda peaks may actually be observed for a wide range of azimuthal angles of transmission. For instance, at normal incidence, and when unpolarized incident light is transmitted through the surface into s-polarized light, the Yoneda peaks occur around $\theta_t=\theta_t^\star$ [or $\qp=\sqrt{\e_1}\omega/c$] independent of the value of the azimuthal angle of transmission $\phi_t$, and they will have constant height~[Fig.~\ref{fig:2Dmdtc_vtp_theta_0}(i)].
Similarly, when unpolarized light is transmitted into p-polarized light for the same scattering system, one observes from Fig.~\ref{fig:2Dmdtc_vtp_theta_0}(h) that a circular groove exists at $\qp=\sqrt{\e_1}\omega/c$. For normal incidence [$\kp=0$], the amplitudes of $\la \partial T_{\ppol\ppol}(\pvec{q} |\pvec{k}) / \partial\Omega_t  \ra_\textrm{incoh}$ and $\la \partial T_{\ppol\spol}(\pvec{q} |\pvec{k}) / \partial\Omega_t  \ra_\textrm{incoh}$ at the position of the groove
will be zero according to \eqref{eq:5.20A} and \eqref{eq:5.20B}. As mentioned earlier, this is due to the factor $\alpha_1(\qp)$, which vanishes when $\qp=\sqrt{\e_1}\omega/c$.

It should be observed from  Figs.~\ref{fig:2Dmdtc_vtp_theta_0}(g)--(i) and \ref{fig:2Dmdtc_ptv_theta_0}(g)--(i), that at normal incidence, and due to the isotropy of the surface, unpolarized incident light will be transmitted by the surface into rotationally symmetric intensity distributions independent of whether the transmitted light is p- or s-polarized.
When unpolarized light is incident from the dielectric, there are only minor differences in the intensity distributions of the p- and s-polarized transmitted light~[Figs.~\ref{fig:2Dmdtc_ptv_theta_0}(h)--(i)]. However, when the light is incident from vacuum, Figs.~\ref{fig:2Dmdtc_vtp_theta_0}(h)--(i) show pronounced differences in their intensity distributions.

\subsection{Non-normal incidence}

%
\begin{figure*}
  \centering
  \includegraphics[width=0.47\textwidth]{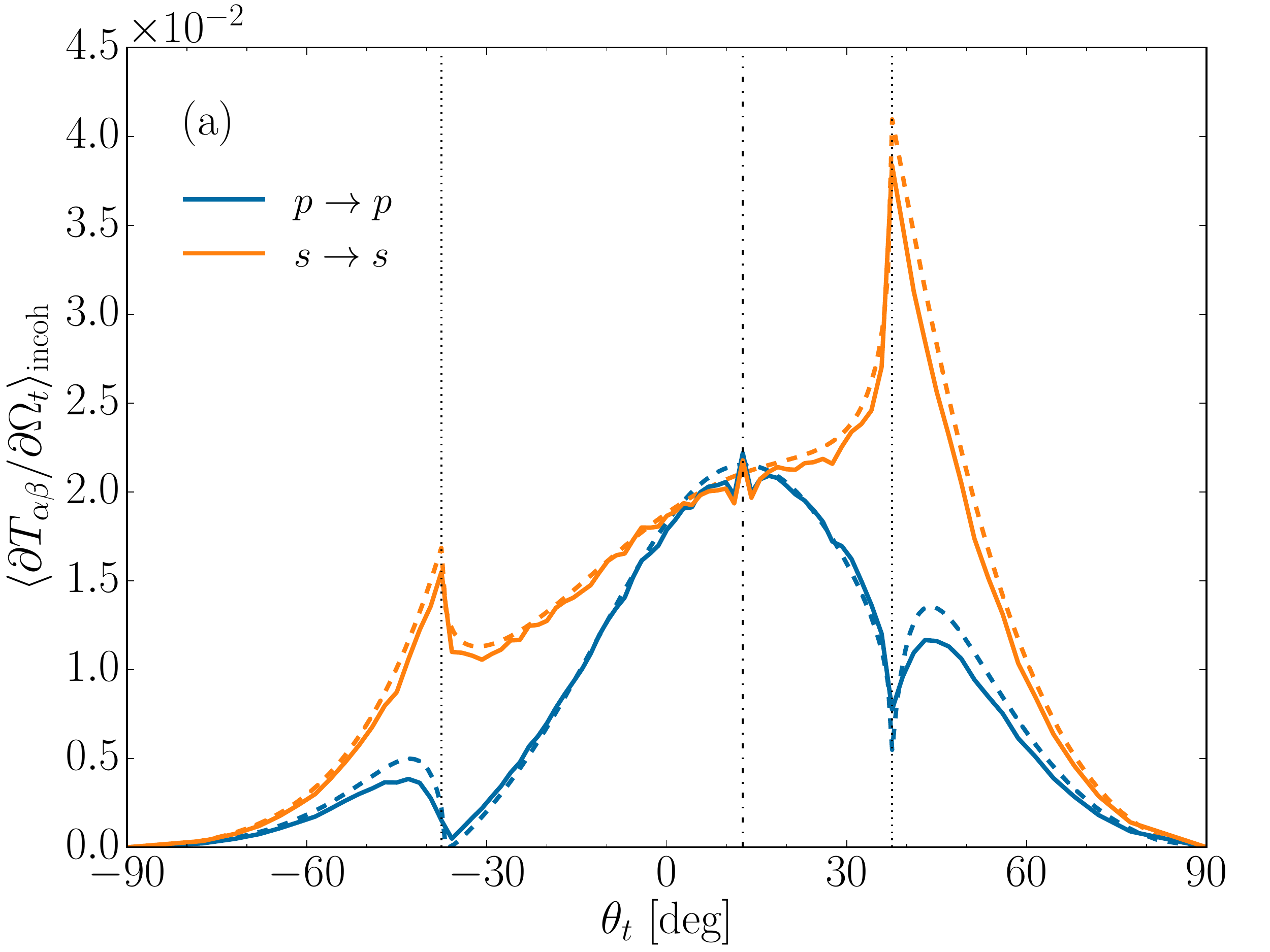}
  \includegraphics[width=0.47\textwidth]{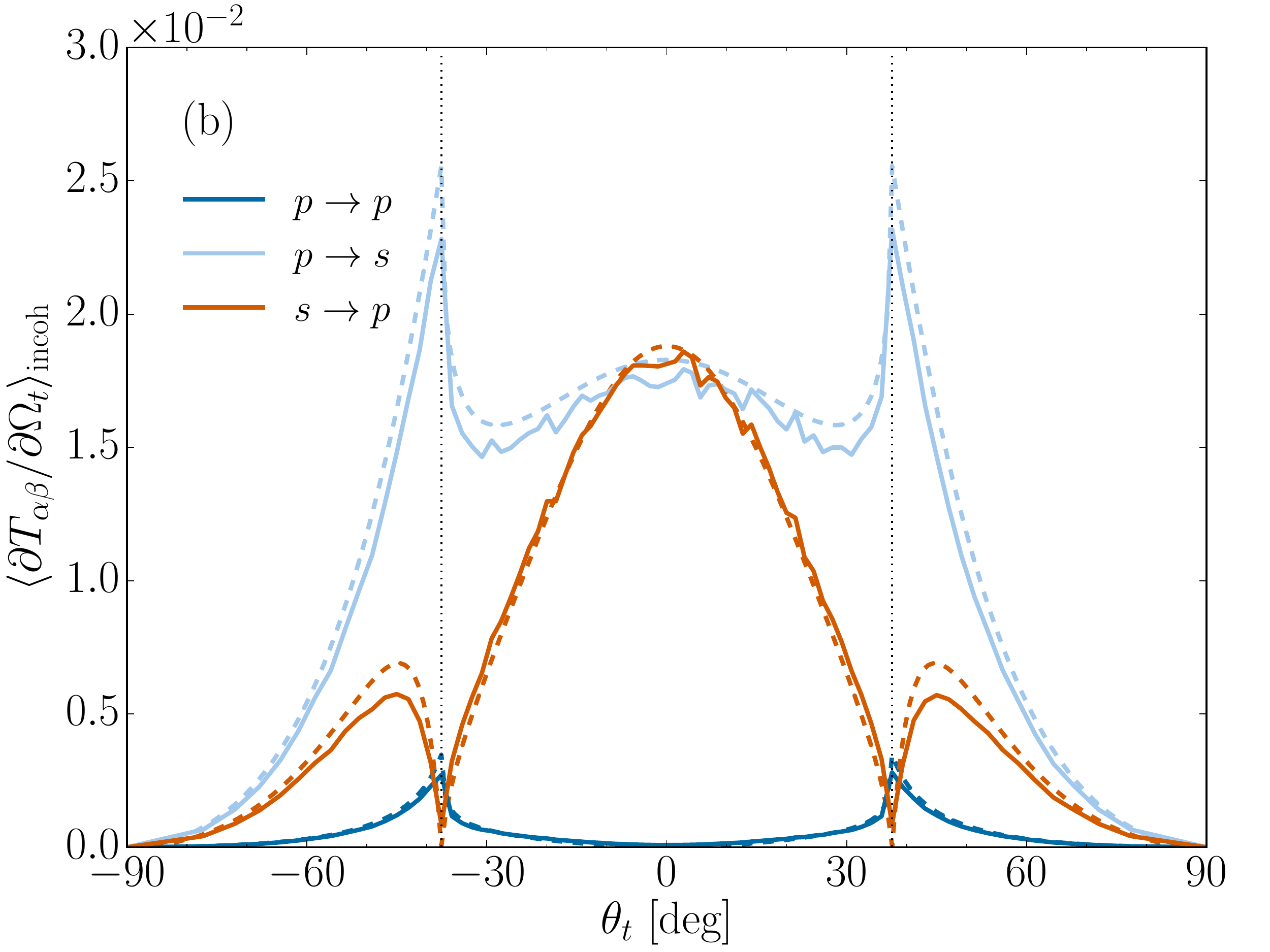}
  \caption{
    (a) Same as Fig.~\ref{fig:inplane_mdtc_theta_0}(a) but for angles of incidence $(\theta_0,\phi_0)=(\ang{21.1},\ang{0})$.
   (b) Same as Fig.~\ref{fig:cut_mdtc_vtp-20}(a) but for out-of-plane scattering [$\phi_t=\pm\ang{90}$].
   Results for combinations of the polarizations of the incident and scattered light  for which the scattered intensity was everywhere negligible have been omitted. [Parameters: $\e_1=1.0$, $\e_2=2.6896$; $\delta=\lambda/20$, $a=\lambda/4$].
 }
\label{fig:cut_mdtc_vtp-20}
\end{figure*}
%
\begin{figure*}
  \centering
  \includegraphics[width=0.8\textwidth]{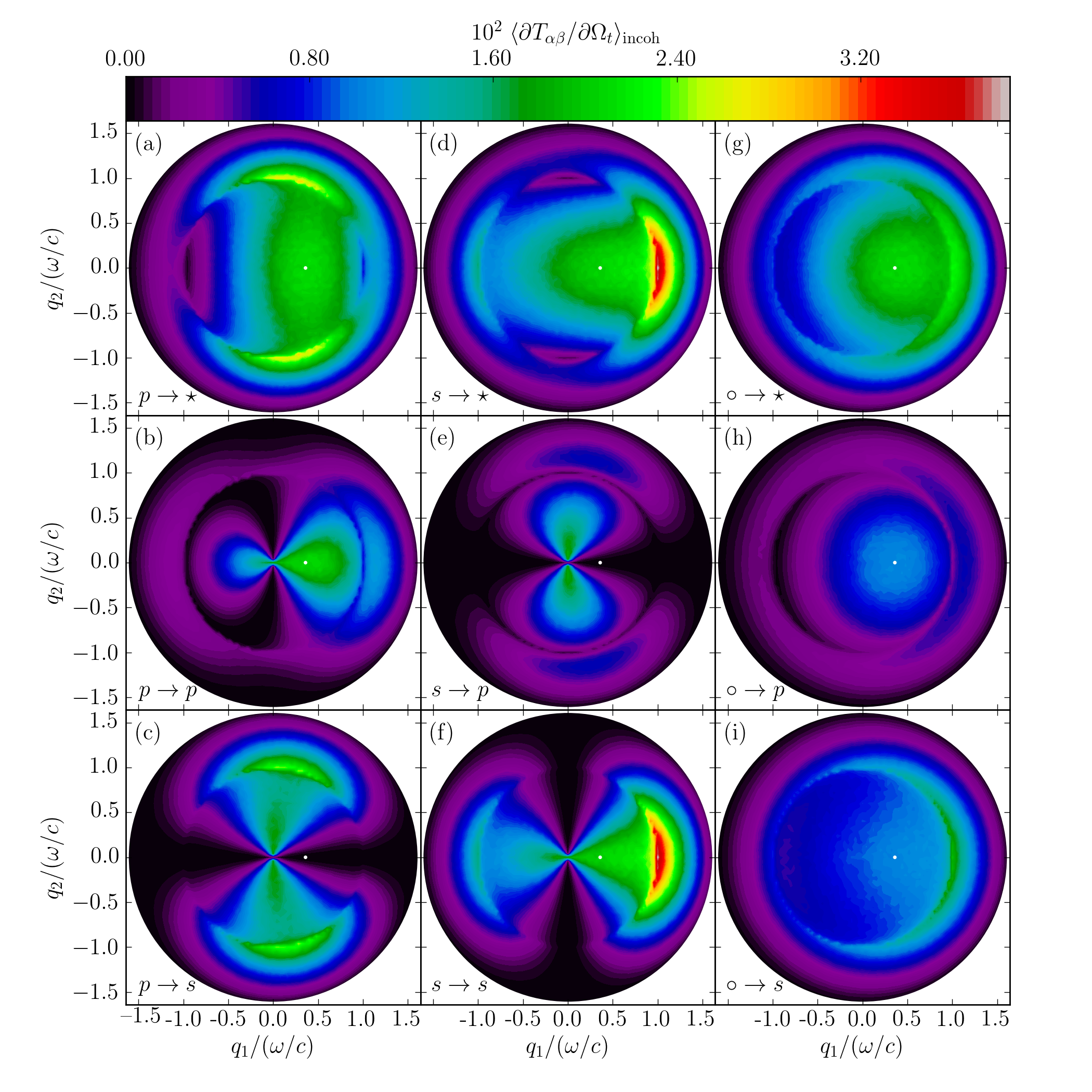}
  \caption{
  Same as Fig.~\protect\ref{fig:2Dmdtc_vtp_theta_0}, but for the angles of incidence $(\theta_0,\phi_0)=(\ang{21.1},\ang{0})$.
  }
  \label{fig:vtp_2D-20}
\end{figure*}
%

We now address the situation when $\theta_0\neq \ang{0}$, and we start our discussion by assuming that the light is incident from vacuum onto its rough interface with the dielectric.
In Fig.~\ref{fig:cut_mdtc_vtp-20} we present the MDTC for light that has been transmitted incoherently (a) in-plane and (b) out-of-plane by the surface for $\theta_0=\ang{21.1}$, and in Fig.~\ref{fig:vtp_2D-20} we present the corresponding full angular intensity distributions.
Figures~\ref{fig:cut_mdtc_vtp-20} and \ref{fig:vtp_2D-20} show that the Yoneda peaks are still prominent, but their amplitudes are no longer independent of the azimuthal angle of transmission, as was found for normal incidence.
For $\spol\to\spol$ transmission, Figs.~\ref{fig:cut_mdtc_vtp-20}(a) and \ref{fig:vtp_2D-20}(f), it is found that the Yoneda peak amplitudes are higher in the forward transmission plane than in the backward plane, and the former peaks have a higher amplitude than they had for  normal incidence.
Moreover, the Yoneda peaks visible in cross-polarized $\ppol\to\spol$ transmission, Fig.~\ref{fig:vtp_2D-20}(c), that for normal incidence were located symmetrically out-of-plane, are now moving into the forward transmission plane. 
The amplitude of $\T{\ppol\alpha}{q}$ when $\qp=\sqrt{\e_1}\w/c$, which was essentially zero for normal incidence, no longer vanishes everywhere as can be seen in Fig.~\ref{fig:cut_mdtc_vtp-20} and the second row of subfigures in Fig.~\ref{fig:vtp_2D-20}, but we do still observe a local minimum in the transmitted intensity into p-polarized light at the position of the Yoneda peaks, and this intensity is, in the plane of incidence, substantially lower than the corresponding intensity for transmission into s-polarized light.

\begin{figure}
  \centering
    \includegraphics[width=0.5\textwidth]{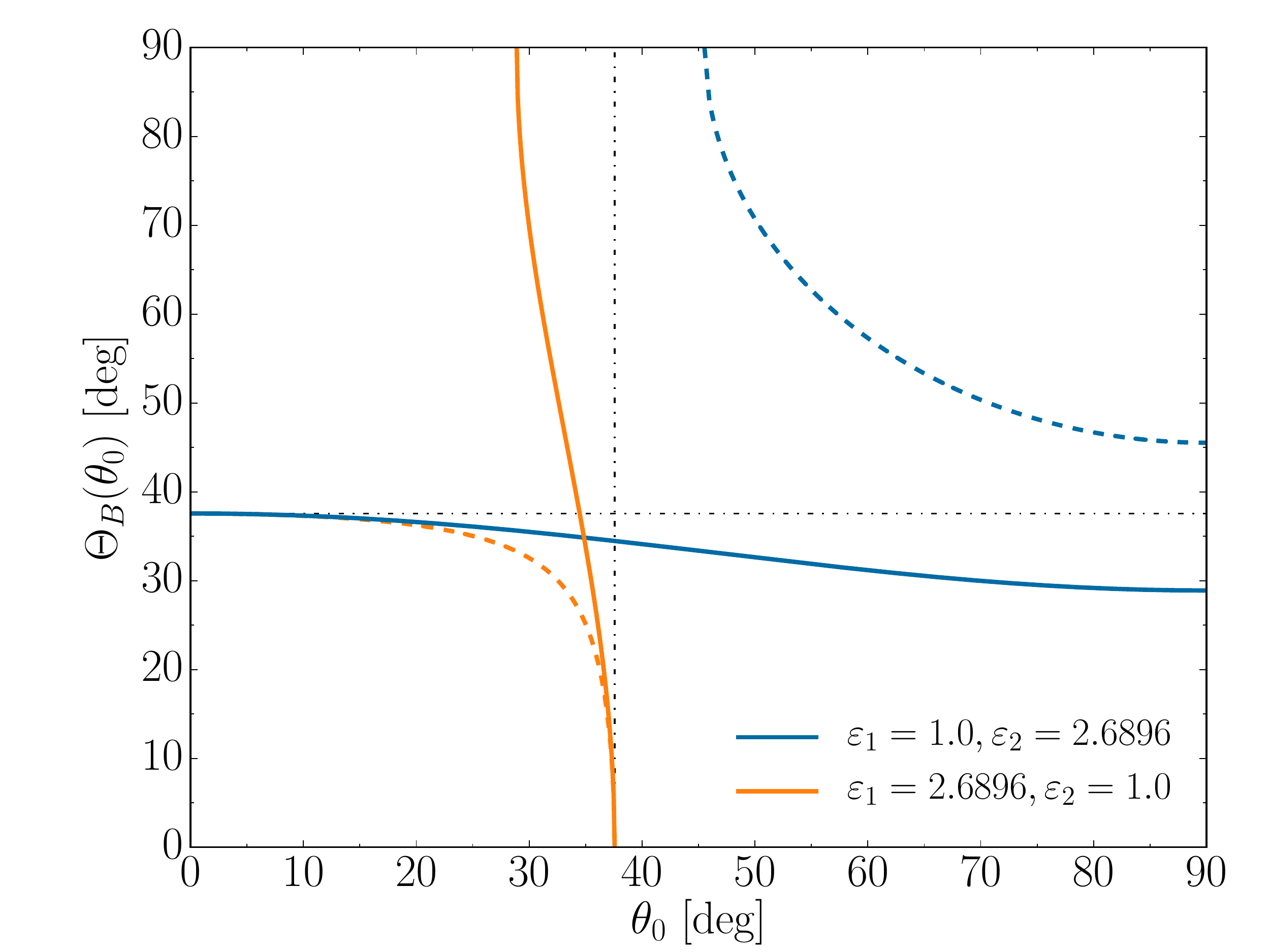}
    \caption{
    Dependence of the in-plane Brewster scattering angle $\Theta_B$ on the polar angle of incidence $\theta_0$ for $\phi_t=\ang{180}$ [Eq.~\eqref{eq:B}]. Corresponding results, but for $\Theta_B$ in reflection and $\phi_s=\ang{0}$ as provided by Eq.~(56) in Ref.~\citenum{Hetland2016a}, are included as dashed lines for completeness.
    The critical angle $\theta_t^\star$ has been indicated on both axes as black dash-dotted lines.
    }
  \label{fig:B}
\end{figure}
Further inspection of Fig.~\ref{fig:cut_mdtc_vtp-20} for $\ppol\to\ppol$ co-polarized transmission reveals that the local minimum found in-plane in the backscattering direction ($\phi_t=\ang{180}$), has shifted its position away from the critical polar angle of $\theta_t^\star$. To first order in SAPT, for which the transmitted intensity at this local minimum is zero, this shift is due to behaviour in the function $F(\bqp|\bkp)$ [Eq.~\eqref{eq:F}] that deserves a more thorough discussion.
When $\kp\neq 0$, $F(\bqp|\bkp)$ can only cause $\T{pp}{q}$ to vanish for $\pvec{q}\cdot\pvec{k}<0$ (backward scattering). Specifically, for in-plane backward scattering [$\pvecUnit{q}\cdot\pvecUnit{k}=-1$], $\T{pp}{q}$ will be zero for angles of transmission
\begin{align}
  \Theta_B(\theta_0) = \sin^{-1}
  \left(
    \frac{\e_1}{\e_2}
    \sqrt{\frac{\e_2}{\e_1}-\sin^2{\theta_0}}
  \right).
   \label{eq:B}
\end{align}
Note that for normal incidence, $\theta_0=\ang{0}$, Eq.~\eqref{eq:B} reduces to $\Theta_B(\ang{0})=\sin^{-1}{\sqrt{\e_1/\e_2}}$, which becomes $\theta_t^\star$ when $\e_1<\e_2$.
Figure~\ref{fig:B} shows the dependence of $\Theta_B$ on $\theta_0$ for both configurations of the dielectric and vacuum, provided that $\phi_t=\ang{180}$. In this figure, the critical angle $\theta_t^\star$ has been indicated on both axes as black dash-dotted lines.
Corresponding plots of $\Theta_B$ but for incoherent \textit{reflection} from the rough interface (Eq.~(56) in Ref.~\citenum{Hetland2016a}) have been included in the figure as thicker colored dashed lines.
For $\e_1=1.0$, $\e_2=2.6896$ and $\theta_0=\ang{21.1}$, Eq.~\eqref{eq:B} gives $\Theta_B(\ang{21.1})\approx\ang{36.5}$, in good agreement with what we observe in Fig.~\ref{fig:cut_mdtc_vtp-20}.

The transmission angles defined by $\Theta_B$ were first mentioned in the literature by Kawanishi~\etal~\cite{Kawanishi1997}, where the angular values of $\Theta_B$ in both reflection and transmission were explored through a stochastic functional approach for two-dimensional surfaces. They chose to call the angles at which the first order contribution (according to their approach) to $\T{p \alpha}{q}$ vanishes the \textit{Brewster scattering angles}, as a generalization of the Brewster angle (polarizing angle) in reflection for a flat surface.
In what follows, following Kawanishi~\etal, we will refer to the polar angles of transmission in the plane of incidence at which p- and s-polarized light is transmitted diffusely (incoherently) into light of any polarization with zero, or nearly zero, intensity, the Brewster scattering angles.
This is consistent with our previous investigation into the Brewster scattering angles in reflection, as presented in Ref.~\citenum{Hetland2016a}.

The Brewster angle $\theta_B$ is defined by the zero in the reflectivity from a flat surface, for p polarization at the angle of incidence given by $\theta_0 = \theta_B = \tan^{-1}(\sqrt{\e_2/\e_1})$. For one set of $\{\e_1,\e_2\}$, there is hence only one Brewster angle for incidence in a given medium. However, in contrast, we would like to stress the fact that the Brewster scattering angles for $\ppol\to\ppol$ scattering are present for a wide range of angles of incidence, given by Eq.~\eqref{eq:B} for in-plane transmission.

\begin{figure*}
  \centering
  \includegraphics[width=0.47\textwidth]{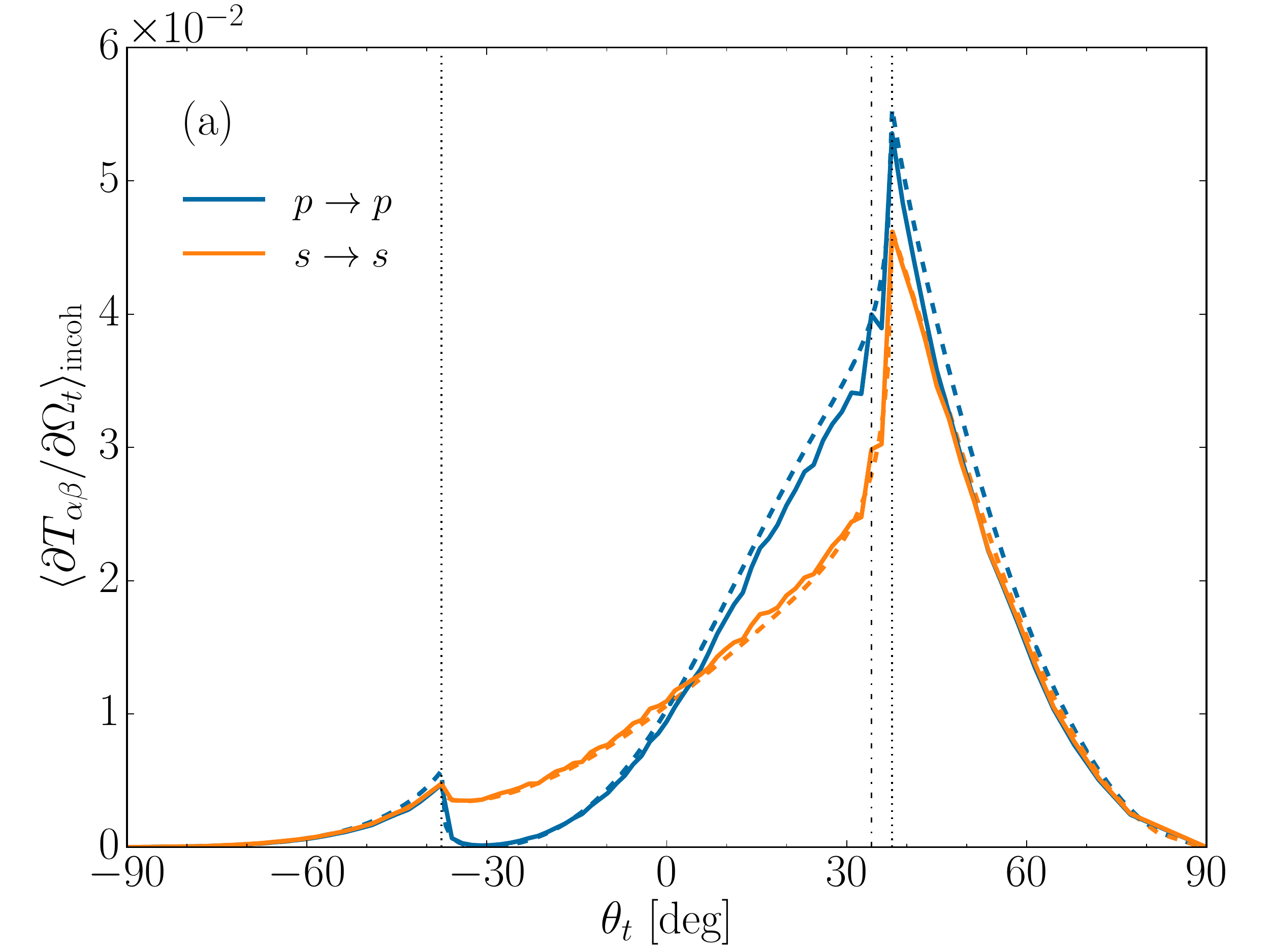}
  \includegraphics[width=0.47\textwidth]{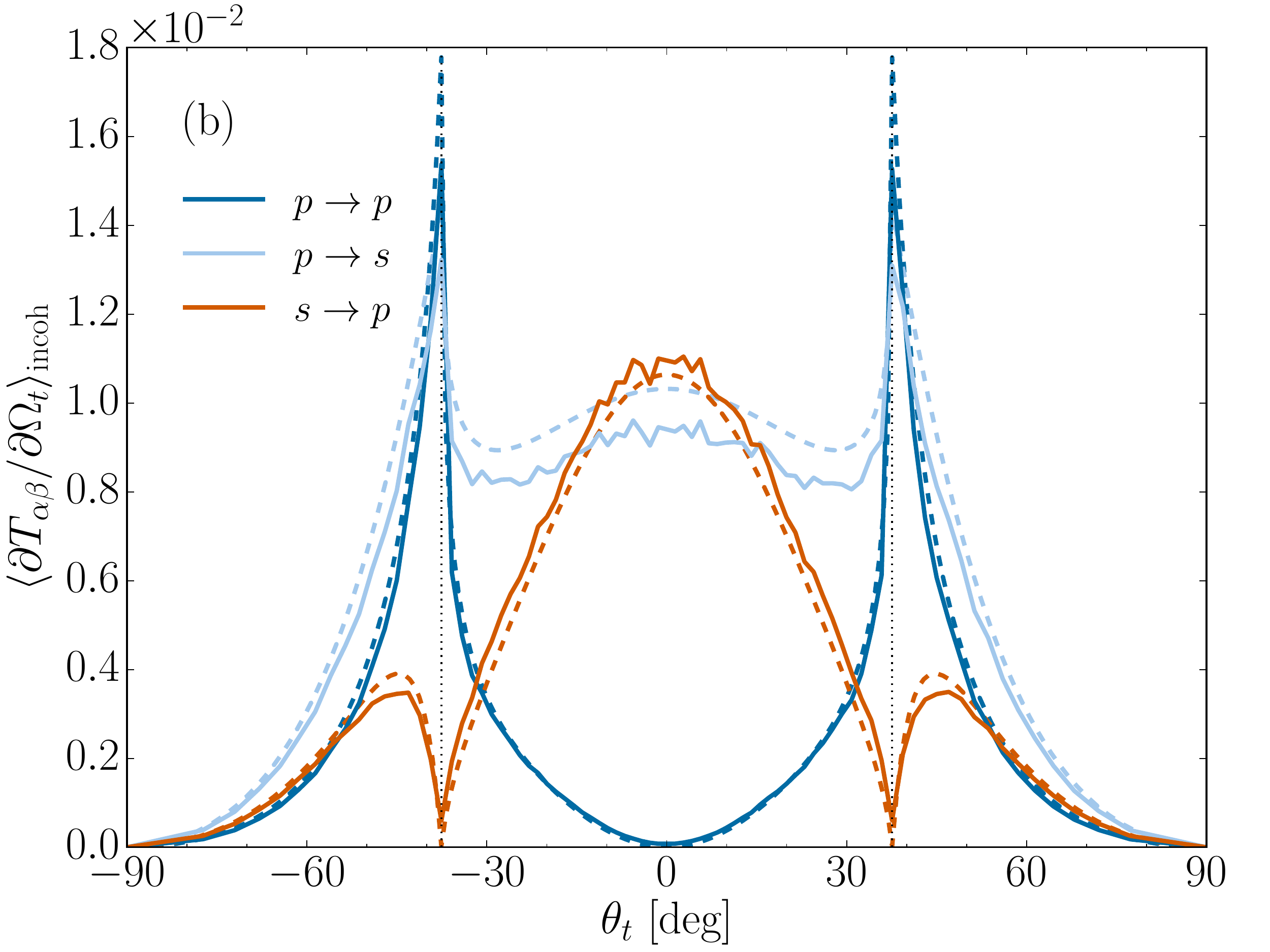}
  \caption{
    (a) Same as Fig.~\ref{fig:inplane_mdtc_theta_0}(a) but for angles of incidence $(\theta_0,\phi_0)=(\ang{66.9},\ang{0})$.
   (b) Same as Fig.~\ref{fig:cut_mdtc_vtp-70}(a) but for out-of-plane scattering [$\phi_t=\pm\ang{90}$].
   Results for combinations of the polarizations of the incident and scattered light  for which the scattered intensity was everywhere negligible have been omitted. [Parameters: $\e_1=1.0$, $\e_2=2.6896$; $\delta=\lambda/20$, $a=\lambda/4$].
 }
\label{fig:cut_mdtc_vtp-70}
\end{figure*}

\begin{figure*}
  \centering
  \includegraphics[width=0.8\textwidth]{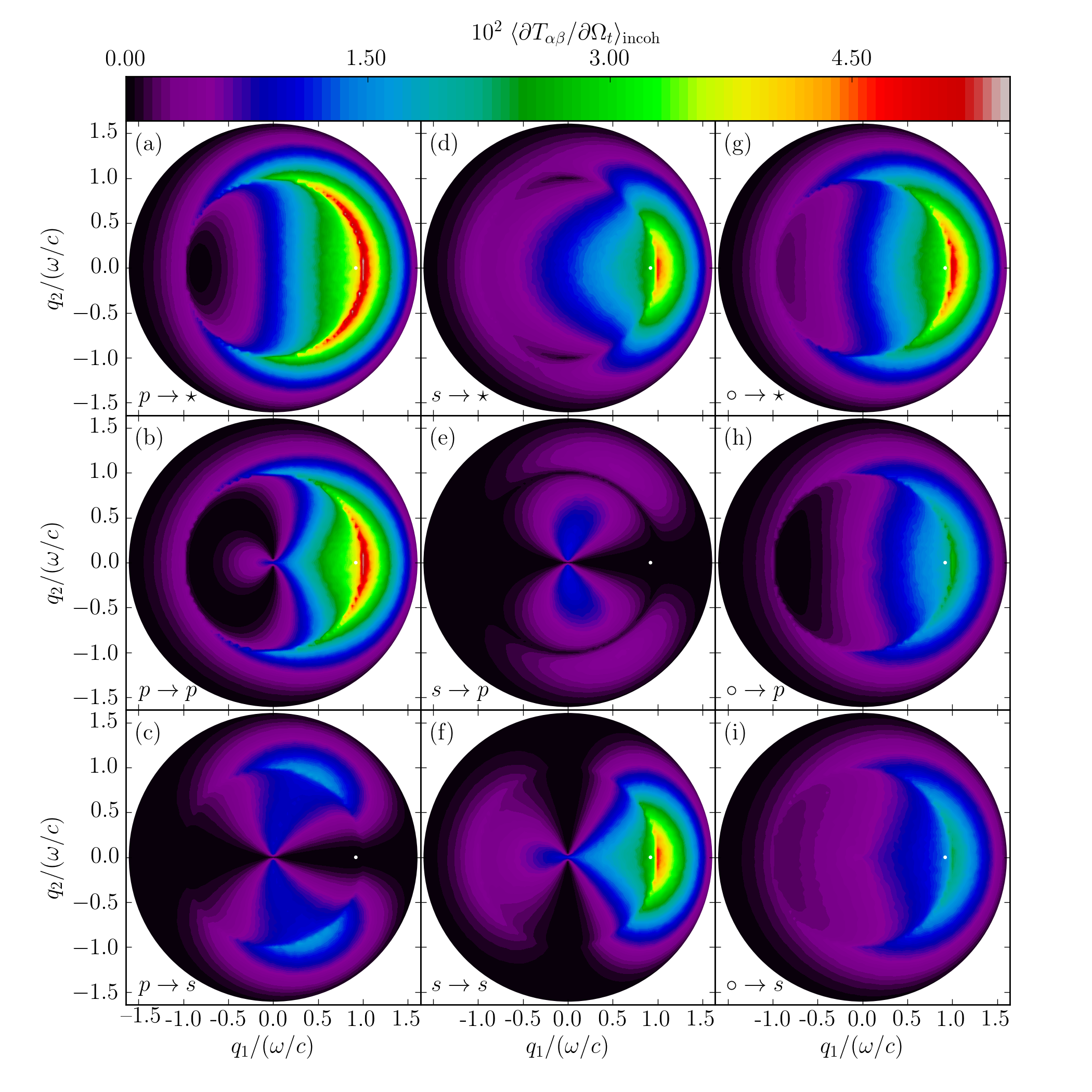}
  \caption{
  Same as Fig.~\protect\ref{fig:2Dmdtc_vtp_theta_0}, but for the angles of incidence $(\theta_0,\phi_0)=(\ang{66.9},\ang{0})$.
  }
  \label{fig:vtp_2D-70}
\end{figure*}

We now let the polar angle of incidence increase to $\theta_0=\ang{66.9}$, as presented in Figs.~\ref{fig:cut_mdtc_vtp-70} and ~\ref{fig:vtp_2D-70}. These figures show that p-polarized transmitted light gives a significant, maybe even dominant, contribution to the in-plane transmitted intensity at the position of the Yoneda peak in the forward transmission plane [$\phi_t=\phi_0$].
This is in sharp contrast to what was found when $\theta_0=\ang{0}$ and  $\theta_0=\ang{21.1}$, where s-polarized transmitted light gave the most significant contribution to the in-plane transmitted intensity at the  position of the Yoneda peaks.
To explain this behavior in the current context, we will again be assisted by Eq.~\eqref{eq:5.20A}, from which it follows that at the position of the Yoneda peaks
\begin{align}
  \left. \left\la \frac{\p T_{pp}(\pvec{q} | \pvec{k} )}{\p\Omega_t}\right\ra_\textrm{incoh} \right|_{\qp=\sqrt{\e_1}\omega/c}
   &\propto
             \frac{\kp^2}{\left|d_p(\kp )\right|^2},
    \label{eq:pp-Yoneda-contrib}
\end{align}
where we used $\alpha_1(\sqrt{\e_1}\omega/c)=0$ in obtaining this result. For normal incidence, Eq.~\eqref{eq:pp-Yoneda-contrib} predicts that the $\ppol\to\ppol$ transmission should go to zero, consistent with what we have seen.
However, as the polar angle of incidence is increased, the function on the right-hand-side of Eq.~\eqref{eq:pp-Yoneda-contrib} will grow quickly, particularly as one approaches grazing incidence.
This has the consequence that $\la \partial T_{\ppol\ppol}(\pvec{q} |\pvec{k}) / \partial\Omega_t  \ra_\textrm{incoh}$, for increasing polar angle of incidence, will go from dipping to peaking at the position of the Yoneda peaks, $\qp=\sqrt{\e_1}\omega/c$. This will not happen for the $\spol\to\ppol$ transmitted light since to lowest order in the surface profile function its intensity is proportional to $\alpha_1(\qp)$, which will always be zero at the position of the Yoneda peaks~[see Eq.~\eqref{eq:5.20C}].

\begin{figure}
  \centering
    \includegraphics[width=1.0\columnwidth]{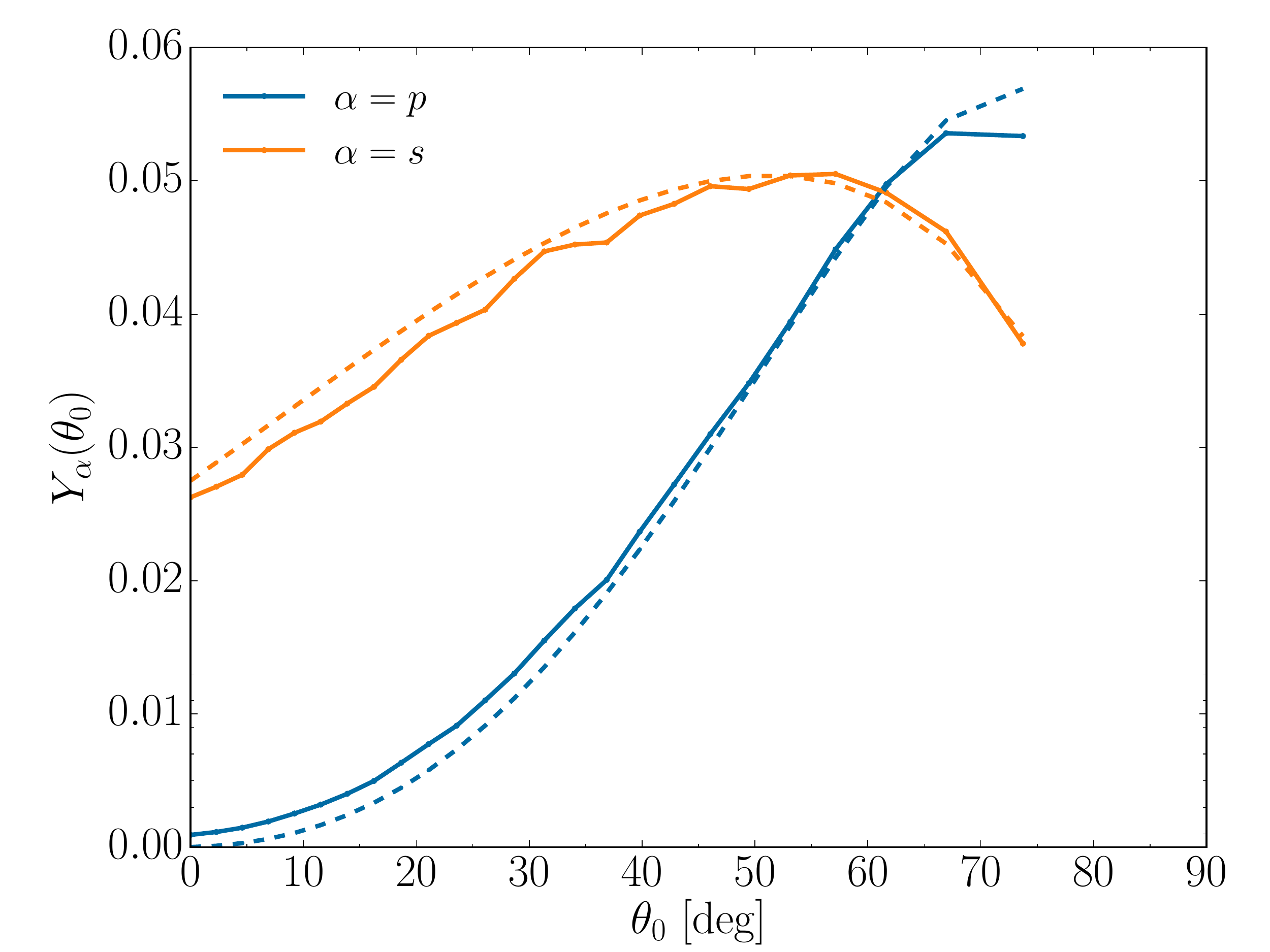}
    \caption{Simulation results for the in-plane, co-polarized contribution to the mean DTC at the Yoneda peak in the forward transmission plane as measured by the function $Y_\alpha(\theta_0)$ defined in  Eq.~\eqref{eq:Y-function}.
    Results for the same angles of incidence, but obtained through SAPT, are included as dashed lines.
    [Parameters: $\e_1=1.0$, $\e_2=2.6896$; $\delta=\lambda/20$, $a=\lambda/4$].
    }
  \label{fig:Yoneda_peak_amp_vs_inc_angle}
\end{figure}

To illustrate this behavior, we study the co-polarized transmitted intensity at the position of the Yoneda peak in the forward transmission plane, $(\theta_t,\phi_t)=(\theta_t^\star,\phi_0)$, by defining the quantity
\begin{align}
  Y_\alpha(\theta_0)
  &\equiv
  \left.
    \left\la
      \frac{\partial T_{\alpha\alpha}(\pvec{q} |\pvec{k})}{ \partial\Omega_t}
    \right\ra_\textrm{incoh}
  \right|_{\pvec{q}=\sqrt{\e_1}\frac{\omega}{c} \pvecUnit{k}}.
  \label{eq:Y-function}
\end{align}
Figure~\ref{fig:Yoneda_peak_amp_vs_inc_angle} presents simulation results for $Y_\alpha(\theta_0)$ as a function of polar angle of incidence for transmission through the vacuum-dielectric system. This figure shows, as is consistent with the preceding discussion, that $Y_\ppol(\theta_0)$ increases more rapidly than  $Y_\spol(\theta_0)$ for moderate angles of incidence; moreover, for an angle of incidence of about \ang{62} and greater, we find that $Y_\ppol(\theta_0)\geq   Y_\spol(\theta_0)$ for the dielectric constants assumed in the current work.
The reason for the nonzero $Y_\ppol(\theta_0=\ang{0})$ is multiple scattering effects which were included consistently in the non-perturbative simulation technique used to obtain the solid-line results of Fig.~\ref{fig:Yoneda_peak_amp_vs_inc_angle}.

Also of interest in the figures presented for $\theta_0=\ang{66.9}$ is the position of the Brewster scattering angle $\Theta_B$, which is now shifted even farther away from the critical angle $\theta_t^\star$. From Eq.~\eqref{eq:B} we calculate that $\Theta_B(\ang{66.9})\approx\ang{30.3}$, in good agreement with the observed value in Fig.~\ref{fig:cut_mdtc_vtp-70}.
This Brewster scattering angle is close to its limiting value for grazing incidence for the dielectric constants currently investigated: $\Theta_B(\ang{90})\approx\ang{28.9}$ [Fig.~\ref{fig:B}].

%
\begin{figure*}
  \centering
  \includegraphics[width=0.47\textwidth]{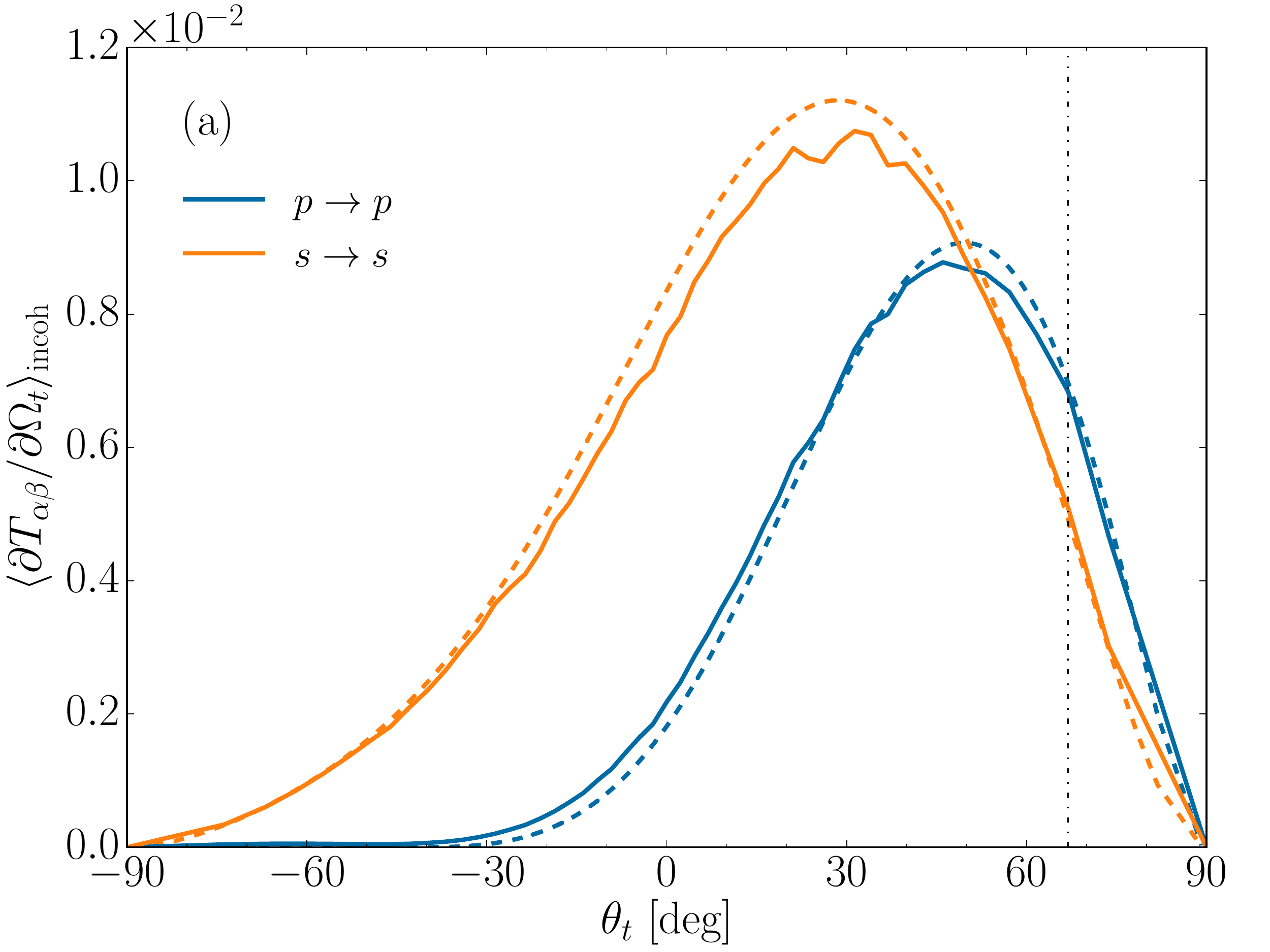}
  \includegraphics[width=0.47\textwidth]{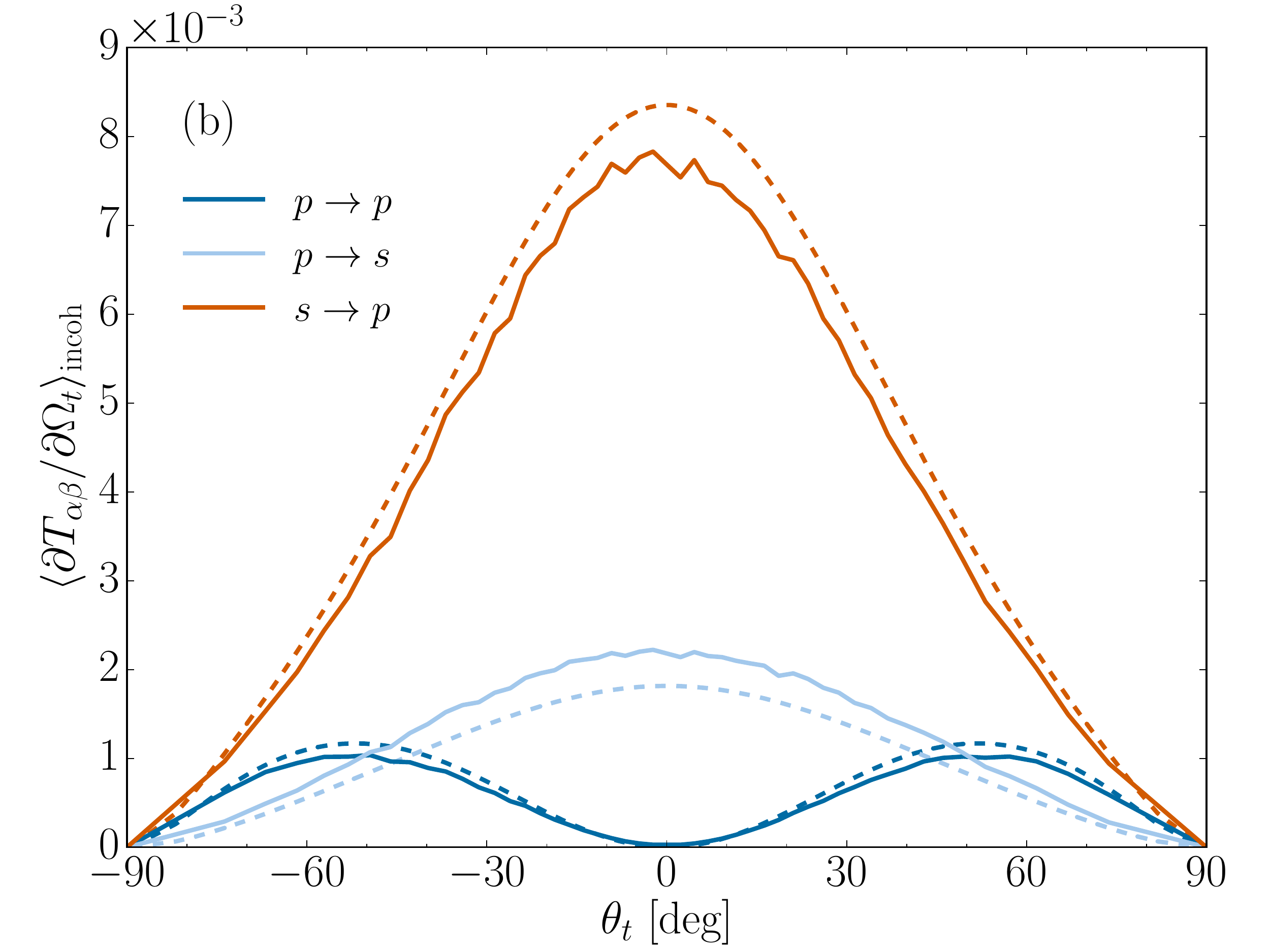}
  \caption{
    (a) Same as Fig.~\ref{fig:inplane_mdtc_theta_0}(b) but for angles of incidence $(\theta_0,\phi_0)=(\ang{34.1},\ang{0})$.
   (b) Same as Fig.~\ref{fig:cut_mdtc-34}(a) but for out-of-plane scattering [$\phi_t=\pm\ang{90}$].
   Results for combinations of the polarizations of the incident and scattered light  for which the scattered intensity was everywhere negligible have been omitted. [Parameters: $\e_1=2.6896$, $\e_2=1.0$; $\delta=\lambda/20$, $a=\lambda/4$].
 }
\label{fig:cut_mdtc-34}
\end{figure*}
%
\begin{figure*}
  \centering
  \includegraphics[width=0.8\textwidth]{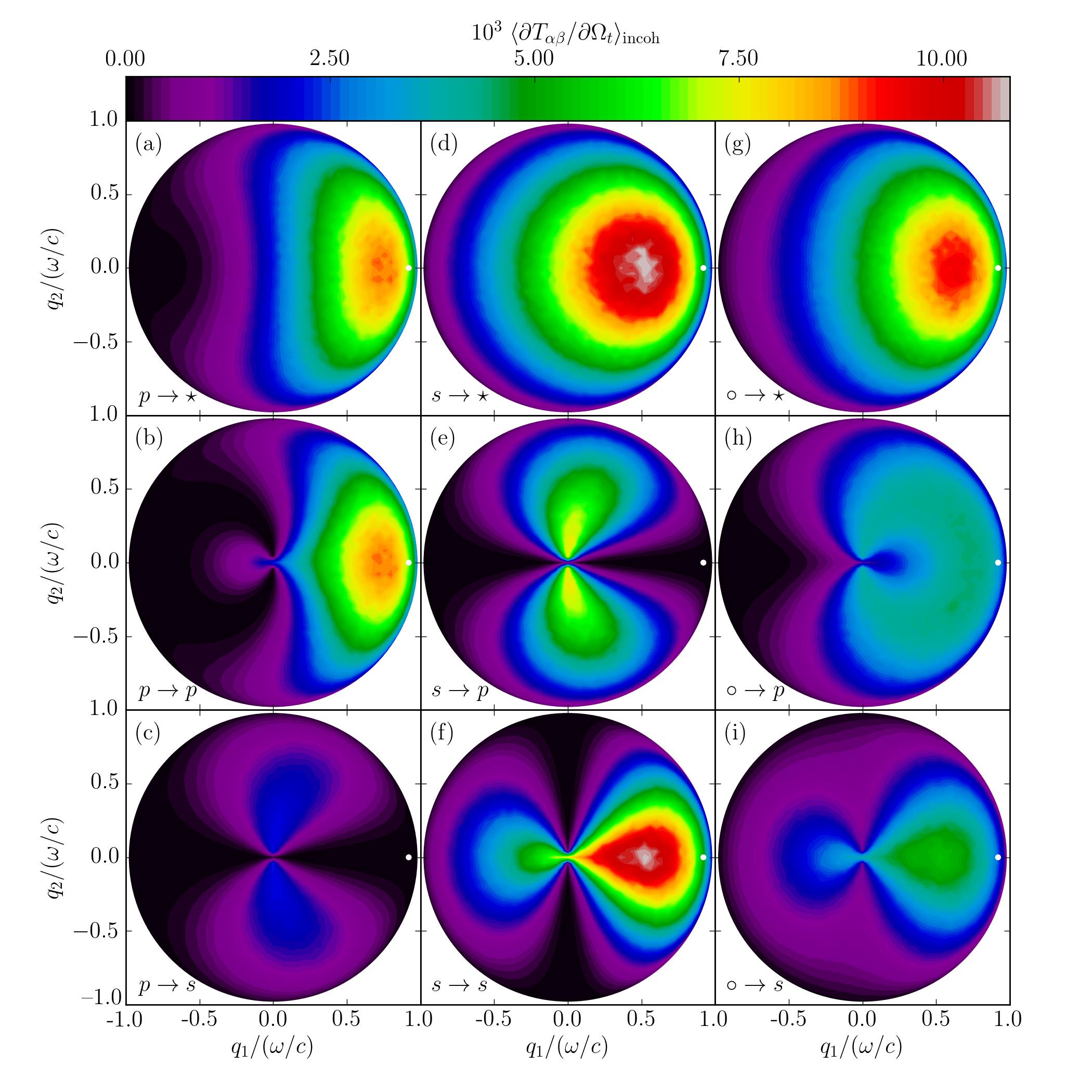}
   \caption{
  Same as Fig.~\protect\ref{fig:2Dmdtc_ptv_theta_0}, but for the angles of incidence $(\theta_0,\phi_0)=(\ang{34.1},\ang{0})$.
  }
  \label{fig:ptv_2D-34}
\end{figure*}
%
\begin{figure*}
  \centering
    \includegraphics[width=0.8\textwidth]{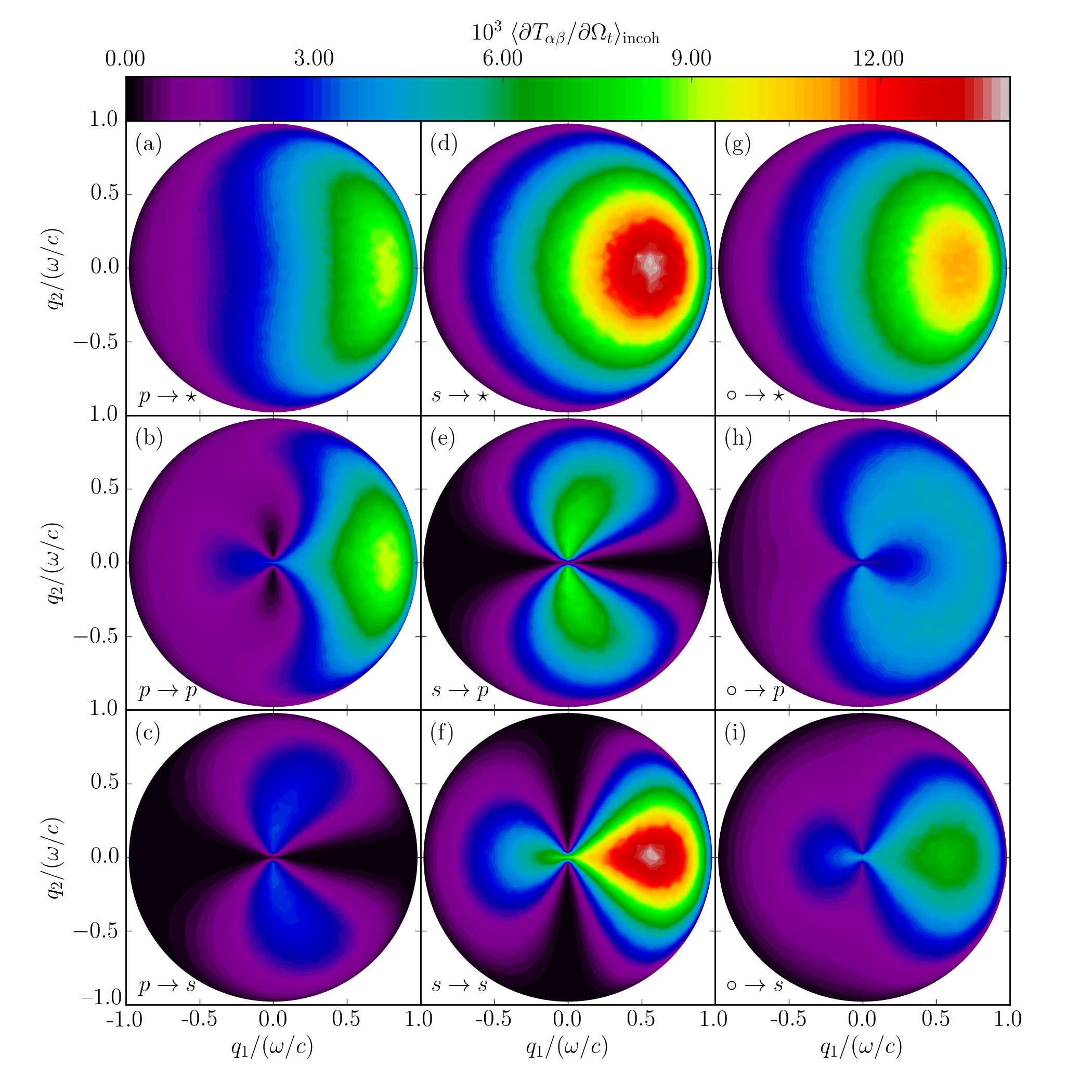}
   \caption{Same as Fig.~\protect\ref{fig:2Dmdtc_ptv_theta_0}, but for the angles of incidence $(\theta_0,\phi_0)=(\ang{45.0},\ang{0})$. Note that for the corresponding flat interface system there would have been zero transmission, since the incident field will experience total internal reflection due to $\theta_0>\theta_t^\star\approx\ang{37.6}$.
   For this reason there is no white dot indicating the specular direction of transmission in this case. For this rough interface system, the light that is transmitted is induced by the surface roughness. }
  \label{fig:ptv_2D-45}
\end{figure*}

We now turn our attention to the inverse system where light is again incident from the dielectric side of the rough interface.
For this system, Fig.~\ref{fig:cut_mdtc-34} presents the (a) in-plane and (b) out-of-plane distributions of the MDTC for a polar angle of incidence $\theta_0=\ang{34.1}$.
As we compare Fig.~\ref{fig:cut_mdtc-34} to Fig.~\ref{fig:inplane_mdtc_theta_0} (b), the observation made for the vacuum-dielectric system that an increase in $\theta_0$ will result in the majority of the light being transmitted into the forward transmission plane, seems also to hold true for the dielectric-vacuum system.
This is expected for weakly rough surfaces like the ones we are investigating, as the main weight of the MDTC to first order in SAPT depends on the power-spectrum factor in Eq.~\eqref{eq:5.20}; a modified gaussian centered at the angular position of the coherently transmitted light.

The Brewster scattering angle can be found also when the light is incident from the dielectric side. For the parameters in Fig.~\ref{fig:cut_mdtc-34}, we find that $\T{pp}{q}$, to first order in SAPT, vanishes at the polar angle of $\Theta_B(\ang{34.1})\approx\ang{40.2}$ for $\phi_t=\ang{180}$. A similar result is presented in the work by Nieto-Vesperinas and S\'{a}nchez-Gil [Fig.~12 in Ref.~\citenum{Nieto-Vesperinas1992}], but the Brewster scattering phenomenon is not mentioned explicitly in this work.

Figures~\ref{fig:ptv_2D-34} and \ref{fig:ptv_2D-45} present the full angular distributions of the MDTC for angles of incidence $(\theta_0,\phi_0)=(\ang{34.1},\ang{0})$ and $(\theta_0,\phi_0)=(\ang{45.0},\ang{0})$, respectively.
The distributions in Figs.~\ref{fig:ptv_2D-34} and \ref{fig:ptv_2D-45} are rather smooth with few, if any, surprising characteristics. It should be noted that the polar angle of incidence $\theta_0=\ang{45.0}$ is larger than the critical angle for total internal reflection, $\theta_0^\star=\sin^{-1}(\sqrt{\e_2/\e_1})\approx\ang{37.6}$, so, for the equivalent planar system, no light would have been transmitted at all; the nonzero intensity distributions observed in Fig.~\ref{fig:ptv_2D-45} are therefore all roughness induced.

\subsection{Transmissivity and transmittance}

\begin{figure*}
  \centering
  \includegraphics[width=0.47\textwidth]{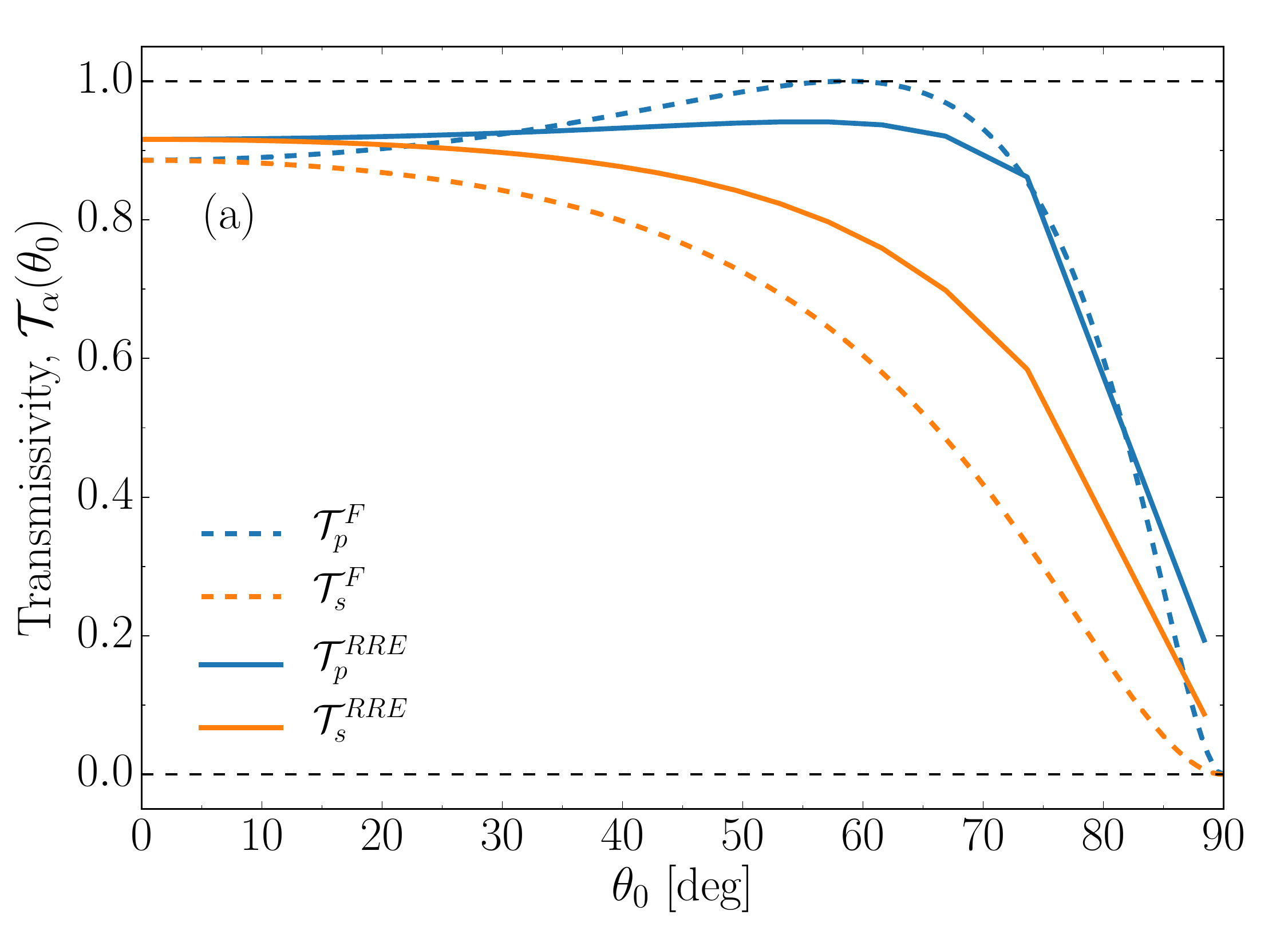}
  \quad
  \includegraphics[width=0.47\textwidth]{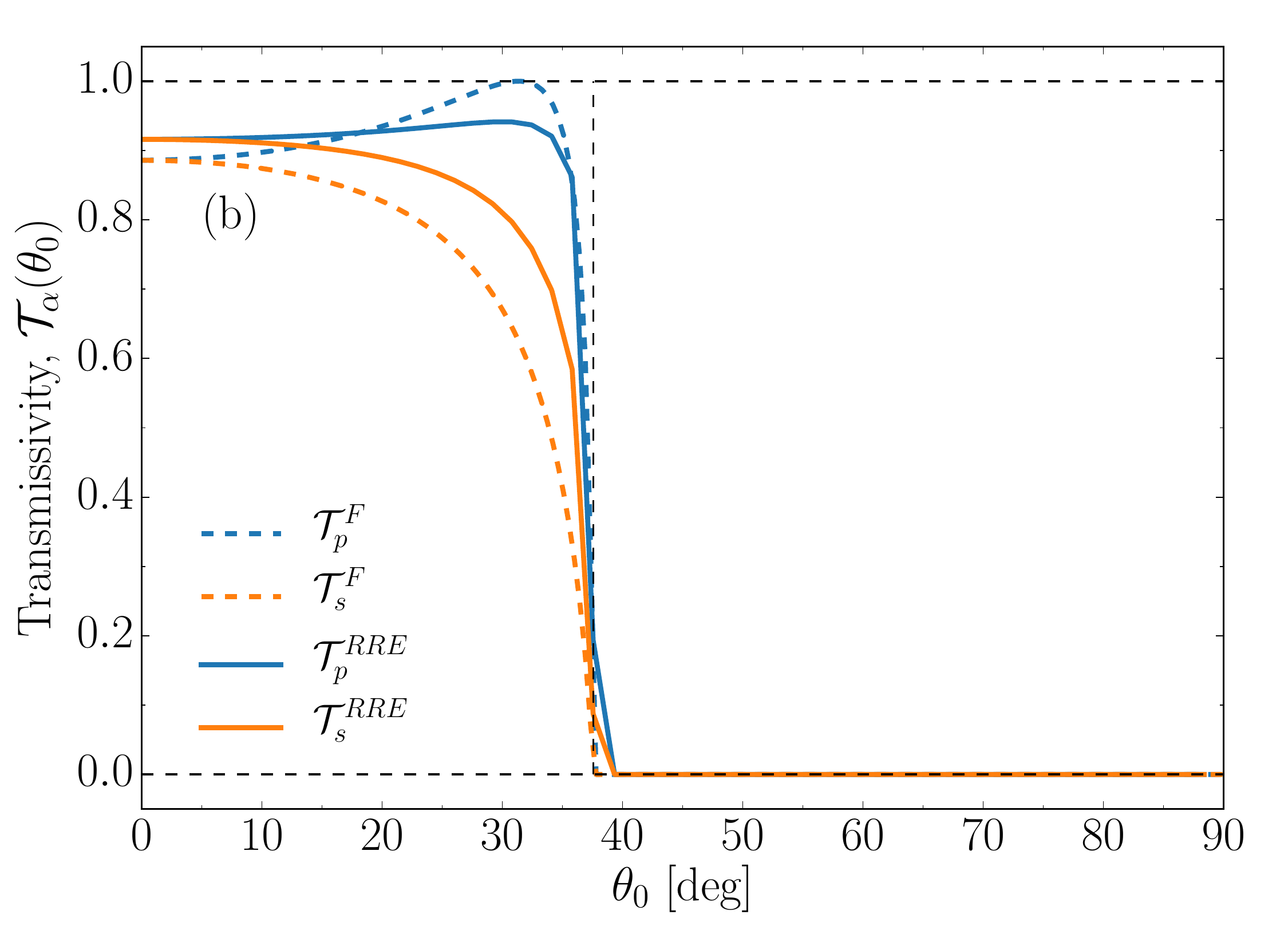}
  \caption{
  (a) The transmissivities ${\mathcal T}_\alpha(\theta_0)$ of a two-dimensional randomly rough vacuum-dielectric interface ($\e_1=1$, $\e_2=2.6896$) for p- and s-polarized light as functions of the polar angle of incidence. (b) The same as in \protect\ref{fig:transmissivity}(a), but for a dielectric-vacuum interface ($\e_1=2.6896$, $\e_2=1$).
  The quantity ${\mathcal T}^F_\alpha(\theta_0)$ indicates the Fresnel transmission coefficient (flat surface transmissivity).
  The critical angle $\theta_0=\theta_0^\star=\sin^{-1}(\sqrt{\e_2/\e_1})$ for total internal reflection for the equivalent planar dielectric-vacuum system is indicated by the vertical dashed line; with the values assumed for the dielectric constants $\theta_0^\star\approx\ang{37.6}$. The roughness parameters assumed in obtaining these results are the same as in Fig.~\ref{fig:inplane_mdtc_theta_0}.
  }
  \label{fig:transmissivity}
\end{figure*}
%
Turning now to the transmissivity (defined in Eq.~\eqref{eq:4.21-new}) of the randomly rough interface, we present in Fig.~\ref{fig:transmissivity}(a) the transmissivity as a function of the polar angle of incidence $\theta_0$ when the interface is illuminated from vacuum by p- and s-polarized light. The transmissivity when the interface is illuminated from the dielectric is presented in Fig.~\ref{fig:transmissivity}(b).
In Fig.~\ref{fig:transmissivity}(a), the transmissivity for incident light of both polarizations is nonzero for all values of $\theta_0$, and tends to zero at a grazing angle of incidence $\theta_0\approx \ang{90}$.
In contrast, the vanishing of the transmissivity for incident light of both polarizations for angles of incidence greater than the critical angle for total internal reflection, $\theta_0^\star=\sin^{-1}(\sqrt{\e_2/\e_1})$, which evaluates to  $\theta_0^\star\approx\ang{37.6}$ for the assumed values of the dielectric constants, is clearly seen in Fig.~\ref{fig:transmissivity}(b).
The transmissivity is larger for p-polarized light than it is for s-polarized light, irrespective of the medium of incidence. This is consistent with the result that the reflectivity of a dielectric surface is larger for s-polarized light than for p-polarized light~\cite{Hetland2016a}.
Even if the transmissivity curves presented in Fig.~\ref{fig:transmissivity} closely resemble the functional form of the transmissivity obtained for equivalent flat interface systems (the Fresnel transmission coefficients, quantified by the dashed lines in Fig.~\ref{fig:transmissivity}), we remark that there are differences.
For instance, from Fig.~\ref{fig:transmissivity} one observes that ${\mathcal T}_\ppol(\theta_0)<1$ for all angles of incidence, while for the equivalent flat interface systems the  transmissivity will be unity at the Brewster angle located around the maxima of ${\mathcal T}_\ppol(\theta_0)$ in Fig.~\ref{fig:transmissivity}.

\begin{figure*}
  \centering
  \includegraphics[width=0.47\textwidth]{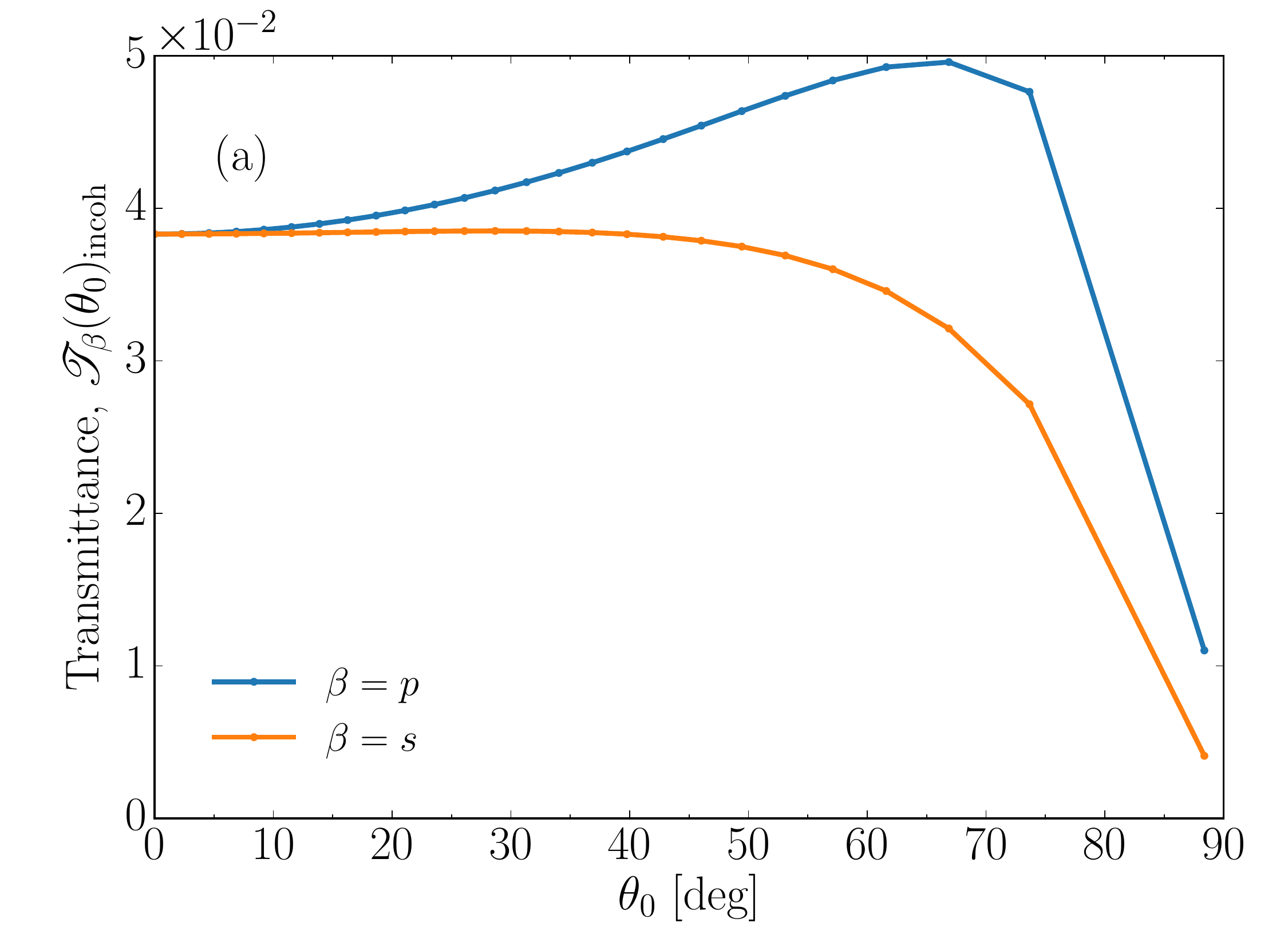}
  \quad
  \includegraphics[width=0.47\textwidth]{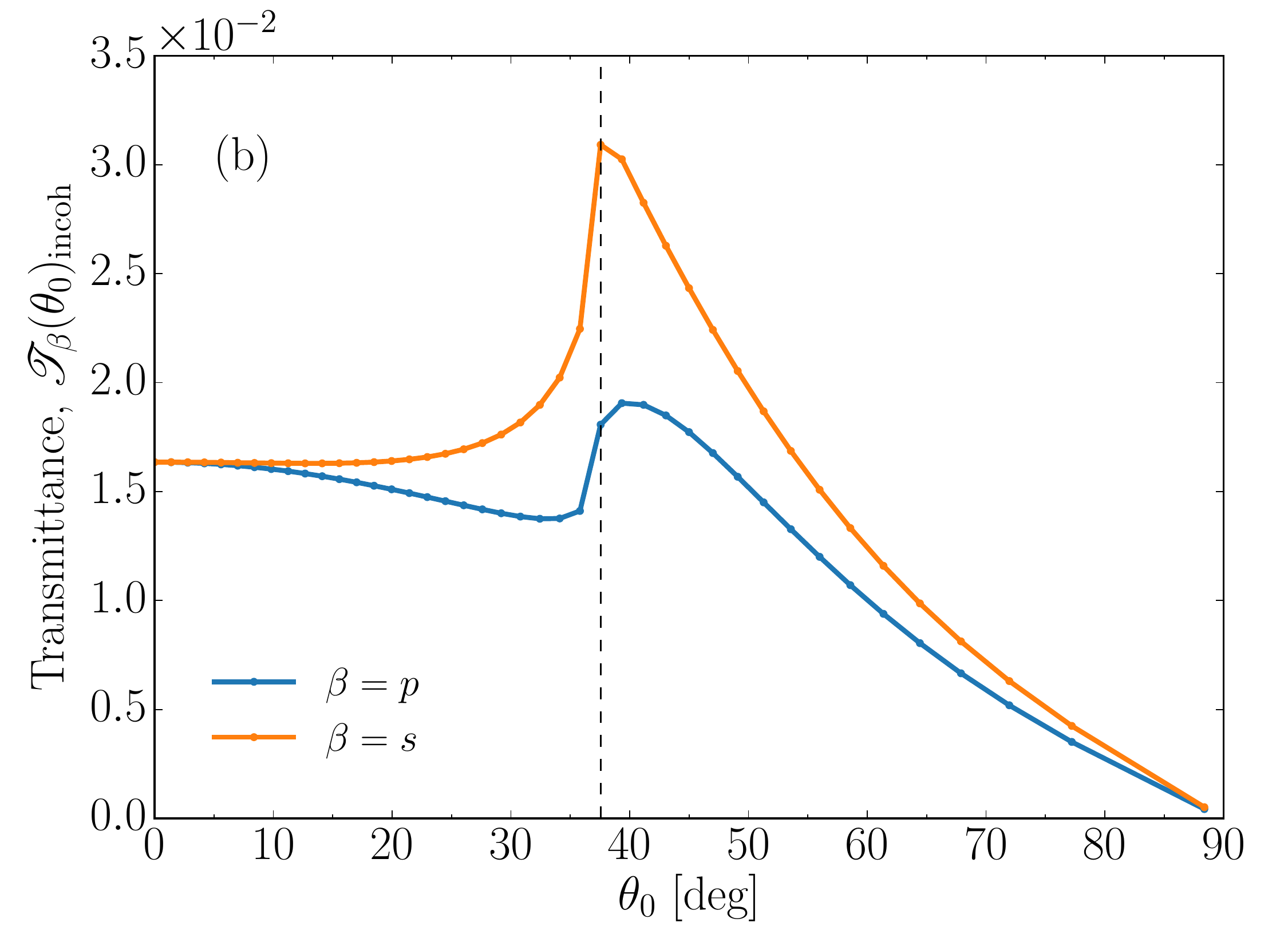}
  \caption{
  The $\theta_0$-dependence of the contribution to the transmittance from p- and s-polarized incident light that has been transmitted incoherently through a two-dimensional randomly rough surface.
  This quantity is for $\beta$-polarized incident light defined by the last term of Eq.~\eqref{eq:transmittance_sum}, \textit{i.e.} ${\mathscr T}_{\beta}(\theta_0)_{\mathrm{incoh}} = {\mathscr T}_{\beta}(\theta_0) - {\mathcal T}_\beta(\theta_0)$.
  The scattering systems assumed in obtaining these results were; (a) vacuum-dielectric  ($\e_1=1$, $\e_2=2.6896$); and (b) dielectric-vacuum ($\e_1=2.6896$, $\e_2=1$).
  The critical angle $\theta_0=\theta_0^\star$ for total internal reflection in the equivalent flat dielectric-vacuum system is indicated by the vertical dashed line. The roughness parameters assumed were the same as in Fig.~\ref{fig:inplane_mdtc_theta_0}. Several simulations were run with small perturbations in the surface length $L$ in order to obtain transmittance data with higher angular resolution (data points are indicated by the solid dots).}
  \label{fig:transmittance_incoh}
\end{figure*}
%

We now focus on the contribution to the transmittance from the light that has been transmitted incoherently through the surface; in Eq.~\eqref{eq:transmittance_sum}, this is the last term denoted by ${\mathscr T}_{\beta}(\theta_0)_{\mathrm{incoh}}$ for incident light of $\beta$ polarization. Small amplitude perturbation theory, through Eq.~\eqref{eq:5.20}, will again assist us in the interpretation of the results.
The transmittance from vacuum into the dielectric is depicted in Fig.~\ref{fig:transmittance_incoh}(a).
In this situation, for which $\e_1<\e_2$, the functions $|d_p(\kp)|^{-2}$ and $|d_s(\kp)|^{-2}$ are both monotonically increasing functions of $\kp$ (or $\theta_0$), and  the transmittances ${\mathscr T}_{\beta}(\theta_0)_{\mathrm{incoh}}$ ($\beta=\ppol,\spol$) are hence slowly varying functions of the angles of incidence, consistent with what is observed in Fig.~\ref{fig:transmittance_incoh}(a).

Figure~\ref{fig:transmittance_incoh}(b) presents the transmittance  ${\mathscr T}_{\beta}(\theta_0)_{\mathrm{incoh}}$ as a function of the polar angle of incidence when the incident medium is the dielectric, and it is found that this quantity displays interesting features.
For instance, in s polarization, a sharp maximum is observed for an angle of incidence a little smaller than \ang{40}, and for this angle of incidence the contribution to the transmittance from the light being transmitted incoherently is about twice the value at normal incidence.
This behavior can be understood on the basis of Eq.~\eqref{eq:5.20D}. As a function of the polar angle of incidence (or $\kp$), the expression for $\la \partial T_{\spol\spol}(\pvec{q} |\pvec{k}) / \partial\Omega_t  \ra_\textrm{incoh}$ in this equation will have a maximum when $|d_s(\kp )|^{-2}$ is peaking.
This happens when $\kp=\sqrt{\e_2}\omega/c$, or equivalently, when $\theta_0=\theta_0^\star$. The expression for the $\spol\to\ppol$ cross-polarized MDTC will also go through a maximum at the same critical angle [see Eq.~\eqref{eq:5.20B}], and so, therefore, will ${\mathscr T}_{\spol}(\theta_0)_{\mathrm{incoh}}$.
This explains the functional dependence of ${\mathscr T}_{\spol}(\theta_0)_{\mathrm{incoh}}$ on the angle of incidence. From Fig.~\ref{fig:transmittance_incoh}(b) it is also observed that the two curves behave differently around $\theta_0=\theta_0^\star$. While the transmittance ${\mathscr T}_{\spol}(\theta_0)_{\mathrm{incoh}}$ is monotonically increasing in the interval $\ang{0}<\theta_0<\theta_0^\star$ and monotonically decreasing in the interval $\theta_0^\star<\theta_0<\ang{90}$, this is not the case for the transmittance of p-polarized incident light.
Similar to the case of s-polarized incident light, the rapid dependence on the angle of incidence of ${\mathscr T}_{\ppol}(\theta_0)_{\mathrm{incoh}}$ around $\theta_0=\theta_0^\star$ is due to the factor $|d_\ppol(\kp )|^{-2}$ present in Eqs.~\eqref{eq:5.20A} and ~\eqref{eq:5.20C}.
However, unlike in the case of s-polarized incident light, the cross-polarized contribution to the MDTC,  $\la \partial T_{\spol\ppol}(\pvec{q} |\pvec{k}) / \partial\Omega_t  \ra_\textrm{incoh}$, Eq.~\eqref{eq:5.20C}, will go to zero at the critical angle $\theta_0=\theta_0^\star$ due to the factor $\alpha_2(\kp)$ that is present in the expression for it.
Therefore, for p-polarized incident light, the transmittance will have a contribution from co-polarized transmission which peaks at the critical angle of incidence, and a contribution from cross-polarization that has a dip down to zero at the critical angle, and it is the sum of the two that results in the functional form observed in ~Fig.~\ref{fig:transmittance_incoh}(b).

\section{Conclusions}

In the current work we have investigated the transmission of light through a two-dimensional, randomly rough interface between two semi-infinite dielectric media. A derivation of the reduced Rayleigh equation for the amplitudes of light transmitted both coherently and incoherently was presented together with expressions for the mean differential transmission coefficient, transmissivity and transmittance. The RRE enables a non-perturbative, purely numerical solution of the surface scattering problem, under the Rayleigh hypothesis.
As an example of the numerical implementation of the RRE, the full angular distribution for both co- and cross-polarized incoherent components of the MDTC were reported together with a discussion on the angular dependence of the transmissivity and transmittance, for configurations of vacuum and an absorptionless dielectric separated by a randomly rough interface with a Gaussian power spectrum and correlation function.

Yoneda peaks, peaks in the incoherent MDTC at the critical polar angle in the medium of transmission where the wavenumber in the medium of incidence turns non-propagating, was shown in all cases of transmission into the denser medium.
These peaks are a dominating feature in the distribution of s-polarized diffusely transmitted light for a wide range of azimuthal angles of scattering, but are suppressed for the p-polarized counterpart when the angle of incidence is at, or close to, normal incidence.
The suppression of p-polarized incoherent scattering in-plane in the backscattering direction ($\phi_t=\ang{180}$)  was found to be of special interest, since the angular position of the local scattering minimum in the MDTC was shown to be dependent on the angle of incidence.
This phenomenon, called the ``Brewster scattering angle'' due to its similarity with the flat-surface Brewster angle, was also observed when the medium of incidence was the dielectric. This is consistent with the findings of Kawanishi~\etal~\cite{Kawanishi1997}.
The developement and behaviour of both Yoneda peaks and Brewster scattering angles were investigated over a wide range of angular parameters, and all observed features were explored through small amplitude perturbation theory.

Small amplitude perturbation theory, to lowest order in the surface profile function, was shown to reproduce our numerical results qualitatively to a high degree of accuracy, both through analytical arguments and a numerical implementation of that theory. This leads us to believe that the features presented in the results can be interpreted as single-scattering effects.

The physical origin of the Yoneda peak phenomenon is still not clear, neither from the existing literature on the topic nor from the results obtained in the present detailed study of it.
We have concluded that it is a single-scattering phenomenon. In addition, our results contradict the explanation for the existence of the Yoneda peaks given by Gorodnichev~\etal\cite{Gorodnichev1988}, who argue that the peaks arise from the multiscale roughness of the surface, which requires that the surface height autocorrelation function should be modeled by a sum of Gaussian functions, rather than by just one. In contrast, the numerical results of the present study, as well as the results of first-order small-amplitude perturbation theory, show explicitly that the representation of $W(\xp)$ by a single Gaussian function, Eq.\eqref{eq:2.4}, is sufficient to produce the Yoneda peaks. Therefore, a systematic study of the physical origin of the Yoneda peaks, and their dependence on polarization, will be left for subsequent work.

As an investigation of the quality of the numerical results presented in this paper, unitarity (energy conservation)~\cite{Simonsen2010} was found to be satisfied with an error smaller than $10^{-4}$ when the scattered energies from both reflection and transmission were added, for the roughness parameters and configurations used.

Calculations of the transmission of light through two-dimensional randomly rough surfaces are challenging, and hence they are still often carried out by means of perturbative and approximate methods. Our approach, through the reduced Rayleigh equations, represents a step towards more accurate but still computationally viable solutions of the problem.
This paper complements our previously published work~\cite{Hetland2016a} on the \textit{reflection} of light from a randomly rough dielectric interface.

\begin{acknowledgments}
The authors would like to thank Jean-Philippe Banon for valuable discussions and contributions to the presented work. The research of \O.S.H. and  I.S. was supported in part by The Research Council of Norway Contract No. 216699.
In addition, I.S. acknowledges financial support from the French National Research Agency (ANR) under contract ANR-15-CHIN-0003-01.
This research was supported in part by NTNU and the Norwegian metacenter for High Performance Computing (NOTUR) by the allocation of computer time.
\end{acknowledgments}

\appendix

\section{Evaluation of $\vec{V}( \gamma  | \pvec{Q} )$}
\label{app:Details}

In this appendix we outline the calculation of the vector $\vec{V}(\gamma| \pvec{Q})$ defined by Eq.~\eqref{eq:3.11a-is}. From Eqs.~\eqref{eq:3.11a-is} and \eqref{eq:3.12-is} it follows immediately that
\begin{align}
  V_3( \gamma  | \pvec{Q} ) = I( \gamma |  \pvec{Q} ).
  \label{eq:AA.3}
\end{align}
The remaining two components of $\vec{V}( \gamma  | \pvec{Q} )$ can be obtained by expanding $\exp \left( - \imu \gamma \zeta(\pvec{x}) \right)$ in powers of the surface profile function and integrating the resulting series term-by-term~($\alpha=1,2$)
\begin{align}
  V_\alpha( \gamma  | \pvec{Q} )
  &= - \int \!\dint^2x_\parallel\;
  \exp \left(-\imu \pvec{Q} \cdot \pvec{x}  \right)
  \zeta_\alpha(\pvec{x})
  \exp \left[ - \imu \gamma \zeta(\pvec{x}) \right]
  \nn \\
  &=
  - \int \!\dint^2x_\parallel\;
    \exp \left(-\imu \pvec{Q} \cdot \pvec{x}  \right)
    \zeta_\alpha(\pvec{x})
    \sum_{n=0}^\infty
    \frac{\left(-\imu\gamma\right)^n }{ n! }
    \zeta^n(\pvec{x} )
  \nn \\
  &=
  -
    \sum_{n=0}^\infty
    \frac{\left(-\imu\gamma\right)^n }{ (n+1)! }
    \int \!\dint^2x_\parallel\;
    \exp \left(-\imu \pvec{Q} \cdot \pvec{x}  \right)
    \frac{ \partial \zeta^{n+1}(\pvec{x}) }{ \partial x_\alpha }
  \nn \\
  &=
  - \frac{ \imu }{ \gamma }
    \sum_{m=1}^\infty
    \frac{\left(-\imu\gamma\right)^m }{ m! }
    \int \!\dint^2x_\parallel\;
    \exp \left(-\imu \pvec{Q} \cdot \pvec{x}  \right)
    \frac{ \partial \zeta^{m}(\pvec{x}) }{ \partial x_\alpha }
    . \label{eq:AA.4}
\end{align}
Introducing the  Fourier representation of the $m$th power of the surface profile function,
\begin{align}
  \zeta^m(\pvec{x})
  &=
    \int \!\frac{\dint^2P_\parallel}{ \left(2\pi \right)^2}\;
    \hat{\zeta}^{(m)}(\pvec{P}) \exp \left(\imu \pvec{P} \cdot \pvec{x}  \right), \qquad m\geq 1,
  \label{eq:AA.5-midlabeled}
\end{align}
into Eq.~\eqref{eq:AA.4}, and evaluating the two resulting integrals after changing their order, yields
\begin{align}
  V_\alpha( \gamma  | \pvec{Q} )
  &=
  \frac{ Q_\alpha }{ \gamma }
  \sum_{m=1}^\infty
  \frac{\left(-\imu\gamma\right)^m }{ m! }
  \hat{\zeta}^{(m)}( \pvec{Q} )
  \nn \\
  &=
  \frac{ Q_\alpha }{ \gamma }
   \left[
     \sum_{m=0}^\infty \frac{\left(-\imu\gamma\right)^m }{ m! }
     \hat{\zeta}^{(m)}( \pvec{Q} )
    - (2\pi)^2 \delta( \pvec{Q} )
  \right]
  \nn \\
  &=
   \frac{ I(\gamma | \pvec{Q} ) }{ \gamma }  Q_\alpha
   - \left( 2 \pi \right)^2 \delta\left( \pvec{Q} \right) \frac{ Q_\alpha }{ \gamma }.
    \label{eq:AA.5}
\end{align}
In the last step we have used the result that
\begin{align}
  I(\gamma | \pvec{Q} )
  &=
  \sum_{n=0}^\infty
  \frac{\left(-\imu\gamma\right)^n }{ n! }
  \hat{\zeta}^{(n)}( \pvec{Q} )
    \label{eq:AA.6}
\end{align}
and $\hat{\zeta}^{(0)}(\pvec{Q})=(2\pi)^2\delta\left( \pvec{Q} \right)$. Equation~\eqref{eq:AA.6} follows readily from Eq.~\eqref{eq:3.12-is} by expanding the latter in powers of the surface profile function and integrating the resulting series term-by-term.

By combining Eqs.~\eqref{eq:AA.3} and \eqref{eq:AA.5} we arrive at the final result
\begin{align}
  \vec{V}( \gamma  | \pvec{Q} )
  &=
     \frac{ I(\gamma | \pvec{Q} ) }{ \gamma }
     \left( \pvec{Q} + \gamma \vecUnit{x}_3  \right)
   - \left( 2 \pi \right)^2 \delta\left( \pvec{Q} \right) \frac{ \pvec{Q} }{ \gamma }.
    \label{eq:AA.7}
\end{align}
We note that the last term of Eq.~\eqref{eq:AA.7}, due to the presence  of the factor $\delta\left( \pvec{Q} \right) \pvec{Q}$, will contribute only if $\pvec{Q}=\vec{0}$. Therefore $\gamma$ must also be zero; in all other cases this term will vanish. For this reason, we will refer to the second term of Eq.~\eqref{eq:AA.7} as the singular contribution to $\vec{V}( \gamma | \pvec{Q} )$.

Technically, $\vec{V}( \gamma | \pvec{Q} )$ is a distribution~\cite{Book:Gelfand1964}; for instance, for the special case $\zeta(\pvec{x})=0$ it follows from Eq.~\eqref{eq:3.11-is} that $\vec{V}( \gamma | \pvec{Q} ) = (2\pi)^2 \delta\left( \pvec{Q} \right) \vecUnit{x}_3$ (which is independent of $\gamma$).
As is true for any distribution, it cannot appear alone in a mathematical expression and should therefore not be evaluated for a single argument as if it were an ordinary function; instead a distribution can only be evaluated after being multiplied by some (test) function. This has the consequence that the singular term of $\vec{V}( \gamma | \pvec{Q} )$ may not necessarily lead to a ``real'' singularity when evaluating the distribution. We will indeed see that this is what happens in our case.

\section{Expansion of $T(\pvec{q}|\pvec{k})$ in powers of the surface profile function}
\label{app:SAPT}

In this appendix we outline the derivation of Eq.~\eqref{eq:5.20}. We begin with the expansions
\begin{align}
  \label{app:eq:I-expansion}
  I(\gamma|\pvec{Q}) &= \sum_{n=0}^\infty \frac{\left(-\imu \gamma\right)^n}{n!} \hat{\zeta}^{(n)}(\pvec{Q}),
\end{align}
where
\begin{subequations}
  \label{app:eq:A2}
\begin{align}
  \hat{\zeta}^{(n)}(\pvec{Q}) &= \int \dint^2x_\parallel\, \textrm{e}^{-\imu\pvec{Q}\cdot\pvec{x}} \zeta^n(\pvec{x})
  \\
  \hat{\zeta}^{(0)}(\pvec{Q}) &= \left( 2\pi \right)^2 \delta\left( \pvec{Q} \right),
\end{align}
\end{subequations}
and
\begin{align}
  \label{app:eq:T-expansion}
  \vec{T}(\pvec{q}|\pvec{k}) &=
  2 \alpha_1(\kp)
  \sum_{n=0}^\infty \frac{\left(-\imu \right)^n}{n!} \vec{t}^{(n)}(\pvec{q}|\pvec{k}).
\end{align}
In the last equation the superscript $n$ denotes the order of the corresponding term in powers of $\zeta(\pvec{x})$. When Eqs.~\eqref{app:eq:I-expansion} and \eqref{app:eq:T-expansion} are substituted into Eq.~\eqref{eq:3.24}, the latter becomes
\begin{widetext}
\begin{align}
  \begin{aligned}
  \sum_{m=0}^\infty \sum_{n=0}^m \frac{\left(-\imu \right)^m}{m!}  \binom{m}{n} &
  \int \frac{\dint^2\qp}{(2\pi)^2} \left[-\alpha_1(\pp) + \alpha_2(\qp) \right]^{n-1}
  \hat{\zeta}^{(n)}(\pvec{p}-\pvec{q}) \vec{M}(\pvec{p}|\pvec{q})\, \vec{t}^{(m-n)}(\pvec{q}|\pvec{k})
  \\
  &=
  (2\pi)^2 \delta\left(\pvec{p}-\pvec{k}\right) \frac{1}{\e_2-\e_1} \vec{I}_2.
  \label{app:eq:A4}
\end{aligned}
\end{align}

When we equate terms of zero order in $\zeta(\pvec{x})$ on both sides of this equation we obtain
\begin{align}
  \frac{1}{ -\alpha_1(\pp) + \alpha_2(\pp) }
  \vec{M}(\pvec{p}|\pvec{p})\, \vec{t}^{(0)}(\pvec{p}|\pvec{k})
  =
  (2\pi)^2 \delta\left(\pvec{p}-\pvec{k}\right) \frac{1}{\e_2-\e_1} \vec{I}_2.
  \label{app:eq:zero}
\end{align}
With the aid of the relation
\begin{align}
  \frac{1}{ -\alpha_1(\pp) + \alpha_2(\pp) }
  =
  \frac{ \alpha_1(\pp) + \alpha_2(\pp) }{ (\omega/c)^2 \left( \e_2-\e_1 \right) },
\end{align}
Eq.~\eqref{app:eq:zero} can be rewritten in the form
\begin{align}
  \begin{pmatrix}
    \frac{1}{ \sqrt{\e_1\e_2} } \left[  \e_2\alpha_1(\pvec{p}) + \e_1\alpha_2(\pvec{p})  \right]
    &
    0
    \\
    0
    &
    \alpha_1(\pvec{p}) + \alpha_2(\pvec{p})
  \end{pmatrix}
  \begin{pmatrix}
    t^{(0)}_{\ppol\ppol}(\pvec{p}|\pvec{k})
    &
    t^{(0)}_{\ppol\spol}(\pvec{p}|\pvec{k})
    \\
    t^{(0)}_{\spol\ppol}(\pvec{p}|\pvec{k})
    &
    t^{(0)}_{\spol\spol}(\pvec{p}|\pvec{k})
  \end{pmatrix}
  =
  (2\pi)^2 \delta\left(\pvec{p}-\pvec{k}\right) \vec{I}_2,
\end{align}
from which we obtain
\begin{align}
  \begin{pmatrix}
    t^{(0)}_{\ppol\ppol}(\pvec{q}|\pvec{k})
    &
    t^{(0)}_{\ppol\spol}(\pvec{q}|\pvec{k})
    \\
    t^{(0)}_{\spol\ppol}(\pvec{q}|\pvec{k})
    &
    t^{(0)}_{\spol\spol}(\pvec{q}|\pvec{k})
  \end{pmatrix}
  &= (2\pi)^2 \delta\left(\pvec{q}-\pvec{k}\right)
  \begin{pmatrix}
    \frac{ \sqrt{\e_1\e_2} }{ \e_2\alpha_1(\pvec{k}) + \e_1\alpha_2(\pvec{k}) }
    &
    0
    \\
    0
    &
    \frac{ 1 }{ \alpha_1(\pvec{k}) + \alpha_2(\pvec{k}) }
  \end{pmatrix}.
  \label{app:eq:A8}
\end{align}

For $m\geq 1$, Eq.~\eqref{app:eq:A4} can be written as
\begin{align}
  \begin{aligned}
    \frac{1}{ -\alpha_1(\pp) + \alpha_2(\pp) }
    &
    \vec{M}(\pvec{p}|\pvec{p}) \vec{t}^{(m)}(\pvec{p}|\pvec{k})
    +
    \int \frac{\dint^2\qp}{(2\pi)^2} \left[-\alpha_1(\pp) + \alpha_2(\qp) \right]^{m-1}
    \hat{\zeta}^{(m)}(\pvec{p}-\pvec{q}) \vec{M}(\pvec{p}|\pvec{q})\, \vec{t}^{(0)}(\pvec{q}|\pvec{k})
    \\
    &
    +
    \sum_{n=1}^{m-1} \binom{m}{n}
    \int \frac{\dint^2\qp}{(2\pi)^2} \left[-\alpha_1(\pp) + \alpha_2(\qp) \right]^{n-1}
    \hat{\zeta}^{(n)}(\pvec{p}-\pvec{q}) \vec{M}(\pvec{p}|\pvec{q})\, \vec{t}^{(m-n)}(\pvec{q}|\pvec{k})
    =
    \vec{0}.
  \end{aligned}
  \label{app:eq:A9}
\end{align}
If  we use the result that the matrix $\vec{M}(\pvec{p}|\pvec{p})$ is diagonal and hence easily inverted, and that the matrix $\vec{t}^{(0)}(\pvec{q}|\pvec{k})$ is given by Eq.~\eqref{app:eq:A8}, we can simplify Eq.~\eqref{app:eq:A9} into
\begin{align}
  \begin{aligned}
    \vec{t}^{(m)}(\pvec{p}|\pvec{k})
    =&
    -\left(\e_2-\e_1\right) \left[-\alpha_1(\pp) + \alpha_2(\kp) \right]^{m-1}
    \hat{\zeta}^{(m)}(\pvec{p}-\pvec{k})
    \begin{pmatrix}
      \frac{ \sqrt{\e_1\e_2} }{ d_\ppol(\pp) }
      &
      0
      \\
      0
      &
      \frac{ 1 }{ d_\spol(\pp) }
    \end{pmatrix}
    \begin{pmatrix}
      \frac{ \sqrt{\e_1 \e_2} M_{\ppol\ppol}(\pvec{p}|\pvec{k})}{ d_\ppol(\kp) }
      &
      \frac{                  M_{\ppol\spol}(\pvec{p}|\pvec{k})}{ d_\spol(\kp) }
      \\
      \frac{ \sqrt{\e_1 \e_2} M_{\spol\ppol}(\pvec{p}|\pvec{k})}{ d_\ppol(\kp) }
      &
      \frac{                  M_{\spol\spol}(\pvec{p}|\pvec{k})}{ d_\spol(\kp) }
    \end{pmatrix}
    \\
    &
    - \left(\e_2-\e_1\right)
    \sum_{n=1}^{m-1} \binom{m}{n}
    \int \frac{\dint^2\qp}{(2\pi)^2} \left[-\alpha_1(\pp) + \alpha_2(\qp) \right]^{n-1}
    \hat{\zeta}^{(n)}(\pvec{p}-\pvec{q})
    \\
    & \qquad \qquad \qquad \qquad \qquad \qquad \qquad
    \times
    \begin{pmatrix}
      \frac{ \sqrt{\e_1 \e_2} M_{\ppol\ppol}(\pvec{p}|\pvec{q})}{ d_\ppol(\pp) }
      &
      \frac{ \sqrt{\e_1 \e_2} M_{\ppol\spol}(\pvec{p}|\pvec{q})}{ d_\ppol(\pp) }
      \\
      \frac{                  M_{\spol\ppol}(\pvec{p}|\pvec{q})}{ d_\spol(\pp) }
      &
      \frac{                  M_{\spol\spol}(\pvec{p}|\pvec{q})}{ d_\spol(\pp) }
    \end{pmatrix}
    \vec{t}^{(m-n)}(\pvec{q}|\pvec{k}),
  \end{aligned}
  \label{app:eq:A10}
\end{align}
where
\begin{subequations}
  \label{app:eq:A11}
\begin{align}
  \label{app:eq:A11a}
  d_\ppol(\pp) &=   \e_2 \alpha_1(\pp) + \e_1 \alpha_2(\pp)
  \\
  \label{app:eq:A11b}
  d_\spol(\pp) &=        \alpha_1(\pp) +      \alpha_2(\pp).
\end{align}
\end{subequations}
Equation~\eqref{app:eq:A10} allows $\vec{t}^{(m)}(\pvec{p}|\pvec{k})$ to be obtained  recursively in terms of $\vec{t}^{(m-1)}(\pvec{p}|\pvec{k})$, \ldots, $\vec{t}^{(1)}(\pvec{p}|\pvec{k})$.

When $m=1$, we obtain from Eq.~\eqref{app:eq:A10} the result
\begin{align}
  \vec{t}^{(1)}(\pvec{q}|\pvec{k})
    =&
    -\left(\e_2-\e_1\right) \hat{\zeta}^{(1)}(\pvec{q}-\pvec{k})
    \begin{pmatrix}
      \frac{ \e_1 \e_2        M_{\ppol\ppol}(\pvec{q}|\pvec{k})}{ d_\ppol(\qp) d_\ppol(\kp) }
      &
      \frac{ \sqrt{\e_1 \e_2} M_{\ppol\spol}(\pvec{q}|\pvec{k})}{ d_\ppol(\qp) d_\spol(\kp) }
      \\
      \frac{ \sqrt{\e_1 \e_2} M_{\spol\ppol}(\pvec{q}|\pvec{k})}{ d_\spol(\qp) d_\ppol(\kp) }
      &
      \frac{                  M_{\spol\spol}(\pvec{q}|\pvec{k})}{ d_\spol(\qp) d_\spol(\kp) }
    \end{pmatrix}.
    \label{app:eq:A12}
\end{align}
The matrix elements $\left\{ M_{\alpha\beta}(\pvec{q}|\pvec{k})\right\}$ are given by Eq.~\eqref{eq:3.25a}.

In view of Eq.~\eqref{app:eq:T-expansion} we find that through terms linear in the surface profile function
\begin{align}
  \begin{aligned}
    \vec{T}(\pvec{q}|\pvec{k})
    =&
    (2\pi)^2 \delta\left(\pvec{q}-\pvec{k}\right)
  \begin{pmatrix}
    \frac{ \sqrt{\e_1\e_2} }{ d_\ppol(\kp) }
    &
    0
    \\
    0
    &
    \frac{ 1 }{ d_\spol(\kp) }
  \end{pmatrix}
  2 \alpha_1( \kp)
  \\
  &
  + \imu (\e_2-\e_1) \hat{\zeta}^{(1)}(\pvec{q}-\pvec{k})
    \begin{pmatrix}
      \frac{ \e_1\e_2 M_{\ppol\ppol}(\pvec{q}|\pvec{k}) }{ d_\ppol(\qp) \, d_\ppol(\kp) }
      &
      \frac{ \sqrt{\e_1\e_2} M_{\ppol\spol}(\pvec{q}|\pvec{k}) }{ d_\ppol(\qp) \, d_\spol(\kp) }
      \\ 
      \frac{ \sqrt{\e_1\e_2} M_{\spol\ppol}(\pvec{q}|\pvec{k}) }{ d_\spol(\qp) \, d_\ppol(\kp) }
      &
      \frac{ M_{\spol\spol}(\pvec{q}|\pvec{k}) }{ d_\spol(\qp) \, d_\spol(\kp) }
    \end{pmatrix}
    2 \alpha_1( \kp)
    + \mathcal{O}\left(\zeta^2 \right).
  \end{aligned}
    \label{app:eq:A13}
\end{align}

The substitution of these results into Eq.~\eqref{eq:4.14} and the use of
$\la\hat{\zeta}(\pvec{Q})\hat{\zeta}(\pvec{Q})^*\ra = S\delta^2g(|\pvec{Q}|)$
yields Eq.~\eqref{eq:5.20}.

\end{widetext}

%




%

\end{document}